\documentclass[11pt]{article}
\usepackage{latexsym}
\begin{document}
\thispagestyle{empty}
\begin{flushright}
MPI-Ph/95-52 \\
UNIGRAZ-UTP-13-06-95 \\
June 1995 \\
hep-th/9602027
\end{flushright}
\vskip 15mm
\begin{center}
{\huge{Discussing the U(1)-Problem of $\mbox{QED}_2$ \vskip2mm
without Instantons}${}^{\mbox{\small1}}$} 
\vskip 1cm
\centerline{ {\bf
Christof Gattringer${}^2$ }}
\vskip 15mm
\centerline{Max-Planck-Institut f\"{u}r
Physik, Werner-Heisenberg-Institut}
\centerline{F\"ohringer Ring 6, 80805 Munich, Germany}
\vskip 2mm
\centerline{and}
\vskip 2mm
\centerline{Institut f\"{u}r Theoretische Physik der Universit\"at Graz} 
\centerline{Universit\"atsplatz 5, 8010 Graz, Austria}
\vskip2cm
\end{center}
\begin{center}
{\bf Abstract}
\end{center}
\noindent
We construct $\mbox{QED}_2$ with mass and flavor and an extra Thirring term.
The vacuum expectation values are carefully decomposed into clustering states
using the U(1)-axial symmetry of the considered operators
and a limiting procedure.
The properties of the emerging expectation functional are compared
to the proposed 
$\theta$-vacuum of QCD. The massive theory is bosonized to a generalized
Sine-Gordon model (GSG). The structure of the vacuum of $\mbox{QED}_2$
manifests itself in
symmetry properties of the GSG. We study the U(1)-problem and 
derive a Witten-Veneziano-type formula for the masses of the pseudoscalars 
determined from a semiclassical approximation.
\vskip 1cm
\bigskip \nopagebreak \begin{flushleft} \rule{2 in}{0.03cm}
\\{\footnotesize \ 
${}^1$ accepted for publication in Annals of Physics} \\ 
{\footnotesize \ ${}^2$ e-mail: chg@mppmu.mpg.de} 
\end{flushleft}
\newpage

%
%
\section{Introduction}
There is a long history of attemptes to study problems of four dimensional
field theories in low dimensional models.
Maybe the most prominent example is the U(1)-gauge theory in two 
dimensions first analyzed by Schwinger \cite{schwinger} and therefore
christened Schwinger model. It has many features in common with
QCD. It shows confinement, mass generation of the would-be Goldstone particle 
via the axial anomaly and allows for topologically nontrivial gauge 
field configurations. The model can be made more realistic by introducing 
several flavors and mass terms. The resulting model which we refer to as 
QED$_2$ is less simple to analyze. 
In this paper we study QED$_2$ using the Euclidean functional integral
approach. 

Of course this project is inspired by some `4d mysteries', as 
should be any investigation of toy models. Namely the topics that will 
be attacked are the construction of the $\theta$-vacuum,
the U(1)-problem and Witten-Veneziano type formulas. Those
problems are closely related to each other. 

According to the common wisdom, the $\theta$-vacuum \cite{callan1,callan2} 
is supposed to be the formal superposition of topological sectors
\begin{equation}
| \; \theta \; \rangle \; = \; \sum_{l=-\infty}^{l=+\infty} \; 
e^{- i \theta l} | \; l \; \rangle \; .
\end{equation}
$| l \rangle$ denotes the states formally corresponding to the
sector of classical pure gauges that wind $l$-times around compactified space.
The mathematical status of (1) is of course rather formal, since it is
e.g. unclear how to normalize $| \theta \rangle$. For QED$_2$ a similar,
but also formal double vacuum structure was derived in \cite{lowenstein}.
It has to be remarked, that the explicit form (1) of the $\theta$-vacuum has
dissappeared in later work, in favor of an implicit characterization
of the $\theta$-vacuum through its symmetry properties under 
large gauge transformations (see e.g. \cite{jack2}). 
However for  $\mbox{QED}_2$ we will construct 
the $\theta$-vacuum functional explicitely.

Another rather formal manipulation often is used when one studies 
the contribution of instanton sectors to Euclidean functional integrals. 
Usually expressions like ${\cal D} [ A ]_n$
show up, which are then meant to denote a measure over gauge field 
configurations with fixed winding number $n$. It is indeed a
rather challenging project to marry 
the idea of a winding number which heavily 
relies on the continuity of the classical gauge fields with the concept of
functional measures living on distributions \cite{colella}. For 
QED$_2$ on a torus where the infrared problems are absent,
an interesting construction of the measure with winding number was 
obtained in \cite{joos1}-\cite{wipf6}.

Despite their mathematical problems, the 
topological ideas have played a fruitful role in 
gauge theories and sometimes are a good semiclassical guideline.
In two dimensions where the mathematical analysis of the models is
much simpler there is of course a more elegant strategy. 
In this paper we show that for QED$_2$ it is possible to construct explicitely 
a proper 
clustering vacuum functional 
$\langle .. \rangle^\theta$ 
without relying on instantons or a formal $\theta$-vacuum like (1).
Afterwards it 
is certainly legitimate to compare the properties of 
$\langle .. \rangle^\theta$ with the formal properties of (1).

The second topic that motivies this 2d study is the U(1)-problem
\cite{weinberg}-\cite{crewther2}.
In QCD (as well as in QED$_2$) 
the axial U(1) current $j_5$ acquires an anomaly
when quantizing the theory. In both cases it is possible \cite{bardeen}
to rewrite the right hand side of the anomaly equation as a total 
divergence. Thus one can define a new current $\tilde{j}_5$ which is 
conserved. Ignoring the fact that $\tilde{j}_5$ is not gauge invariant and 
thus rather unphysical, one can formally implement the 
U(1)-axial symmetry related to 
$\tilde{j}_5$. It is known that nature does not respect this
symmetry and thus one could expect a Goldstone particle that corresponds to
this broken symmetry. Since the quarks are massive, one only 
can hope to find an approximate Goldstone particle, i.e. a light
pseudoscalar. Based on the paper by Weinberg \cite{weinberg} for
the case of two flavors it is
believed that the $\eta$-meson which has the right quantum numbers is  
too heavy to play the role of an approximate Goldstone particle. 
For three flavors the corresponding particle is the $\eta^\prime$.
The U(1)-problem now is the absence of a fourth (ninth) light 
pseudoscalar. As will be discussed below, up to some restrictions due 
to the Coleman theorem, the U(1)-problem of QED$_2$ is formulated 
equivalently.

Also for the analysis of the U(1)-problem of QED$_2$ the strategy will
be to study the problem using only  mathematically rigorous
methods, which is of course much easier in two dimensions. The lessons 
on the U(1)-problem will be drawn afterwards and we will discuss
what could be learned for QCD.

Finally we will also analyze Witten-Veneziano-type formulas. They connect
the masses of the pseudoscalar mesons to the topological susceptibility.
It was pointed out in \cite{seilerstam} (see also \cite{diss1,diss2})
that there is a problem
with Witten's original derivation \cite{witten1, witten2}
and an
alternative proof was given. It was argued that the topological 
susceptibility on the
right hand side of the original Witten-Veneziano formula has to be replaced
by its contact term. In a lattice approach \cite{smit}
the formula was generalized to QCD with three flavors of massive
quarks. We show that in $\mbox{QED}_2$ a Witten-Veneziano-type formula for
the masses of the pseudoscalars determined from a semiclassical approximation
holds.

The paper is organized as follows. In the next section we discuss the model 
and its symmetries. In order to treat the mass term perturbatively
we include a Thirring term as an UV-regulator and also a space-time cutoff.
This is followed by a section where we outline the
construction using mass perturbation series.

In Section 4 we show that the vacuum state constructed so far 
does not cluster and thus the vacuum fails to be unique. 
We cure this problem by defining a new vacuum functional 
$\langle .. \rangle^\theta$ in 
a mathematically rigorous way, which
involves the symmetry properties of the operators and a limiting 
procedure. The properties of $\langle .. \rangle^\theta$ give rise
to chiral selection rules which we discuss.

The massive model will be constructed using the bosonization prescriptions
which we establish in Section 5. The bosonic model turns out to be a 
generalized Sine-Gordon model (GSG). Properties of the vacuum functional
$\langle .. \rangle^\theta$ manisfest themselves as symmetry properties 
of the GSG which we discuss. In particular it will turn out that the
symmetry that would correspond to the axial U(1) symmetry of QED$_2$
can not be implemented in the GSG. This leads to the conclusion that
the U(1)-problem of QED$_2$ does not exist.

Using a semiclassical approximation we finally establish that 
a Witten-Veneziano-type formula holds for QED$_2$. 

In Section 6 the relevance and also the limitations 
of the lessons on QED$_2$ for the physics
of QCD will be discussed.

%
%
\section{The model}
The Euclidean action of the model that will be constructed is given by
\begin{equation}
S[\overline{\psi},\psi,A,h] \; = \; S_G[A] + S_h[h] + 
S_F[\overline{\psi},\psi,A,h] + S_M[\overline{\psi},\psi] \; .
\end{equation}
The action for the gauge field reads
\begin{equation}
S_G[A] \; = \; \int d^2x\left( \frac{1}{4} F_{\mu \nu}(x) F_{\mu \nu}(x)
+ \frac{1}{2} \lambda \Big( \partial_\mu A_\mu(x) \Big)^2 \right) \; .
\end{equation}
A gauge fixing term is included that will be considered in the limit 
$\lambda \rightarrow \infty$ which ensures $\partial_\mu A_\mu = 0$
(transverse gauge). As usual $F_{\mu \nu} = 
\partial_\mu A_\nu - \partial_\nu A_\mu$, denotes the field strength
tensor.

\noindent
The fermion action is a sum over N flavor degrees of freedom
\begin{equation}
S_F[\overline{\psi},\psi,A,h] \; = \; \sum_{b=1}^N \int d^2x \;  
\overline{\psi}^{(b)}(x)\gamma_\mu \Big( \partial_\mu - i e A_\mu(x) 
-i g^{1/2} h_\mu(x) \Big)\psi^{(b)}(x) \; .
\end{equation}
For the Euclidean $\gamma$-algebra we choose $\gamma_1 = \sigma_1,
\gamma_2 = \sigma_2, \gamma_5 = \sigma_3$, where $\sigma_i$ are the
Pauli-matrices.
In addition to the gauge field an auxiliary field $h_\mu$ couples to the 
fermions in exactly the same way as $A_\mu$. Its action is given by
\begin{equation}
S_h[h] \; = \; \frac{1}{2}\int d^2x \; h_\mu(x)\Big( \delta_{\mu \nu} - 
\lambda^\prime \partial_\mu \partial_\nu \Big) h_\nu(x) \; .
\end{equation}
$S_h[h]$ is simply a white noise term plus a term that
makes $h_\mu$ transverse in the limit $\lambda^\prime \rightarrow \infty$.

In order to understand the role of the auxiliary field one can
formally integrate over $h_\mu$. This leads to the Thirring term  
\begin{equation}
S_T[\overline{\psi},\psi] = \frac{1}{2} \; g \; \int d^2x \;
\sum_{b=1}^N {j^{(b)}}^T_\mu(x) \sum_{c=1}^N {j^{(c)}}^T_\mu(x) \; ,
\end{equation}
for the transverse part of the U(1)-current $\sum_{b=1}^N {j^{(b)}}^T_\mu$.
The superscript $T$ denotes projection on the transverse direction
\begin{equation}
{j^{(b)}}^T_\mu \; := \; T_{\mu \nu} 
\overline{\psi}^{(b)} \gamma_\nu \psi^{(b)} \; \; \; \; , \; \; \; \; 
T_{\mu \nu} \; := \; \delta_{\mu \nu} - 
\frac{\partial_\mu \partial_\nu}{\triangle} \; ,
\end{equation}
and $T_{\mu \nu}$ is the corresponding projector. The purpose of this Thirring 
term is to make the short distance singularity of
\begin{equation}
\langle \; \overline{\psi}^{(b)}(x) \psi^{(b)}(x) \;
\overline{\psi}^{(b)}(y) \psi^{(b)}(y) \; \rangle \; ,
\end{equation}
integrable. The quoted expression is a typical term showing up in
a power series expansion of the mass term (9) (see below). 
It has to be integrated over
$d^2x d^2y$ which is possible only if an ultraviolet regulator 
such as the Thirring term is
included.

Since the mass term will be treated perturbatively, we denote it separately
\begin{equation}
S_M[\overline{\psi}, \psi] = -\sum_{b=1}^N m^{(b)} \int d^2x \; 
\chi_\Lambda(x) \;
\overline{\psi}^{(b)}(x) \psi^{(b)}(x) \; .
\end{equation}
$m^{(b)}$ are the fermion masses for the various flavors. 
For the perturbation expansion it is necessary to introduce an infrared 
cutoff. Here we use a space-time cutoff, namely a finite rectangle $\Lambda$
in space-time, and $\chi_\Lambda$ denotes its characteristic function.

For vanishing fermion masses $m^{(b)}$, the Lagrangian of 
the model has the symmetry
$\mbox{SU(N)}_L \times \mbox{SU(N)}_R 
\times \mbox{U(1)}_V \times\mbox{U(1)}_A$ 
as is the case for QCD. 
When quantizing the massless theory the axial U(1)-current 
\begin{equation}
j_{5 \; \mu}(x) \; := \; \sum_{b=1}^N \; \overline{\psi}^{(b)} (x)
\gamma_\mu \gamma_5 \psi^{(b)}(x) \; \; ,
\end{equation}
acquires the anomaly
\begin{equation}
\partial_\mu j_{5\; \mu}(x) \; = \; 2 N \; \frac{e}{2( \pi + gN)} \; 
\varepsilon_{\mu \nu} \partial_\mu A_\nu(x) \; 
+  \; \mbox{contact terms} \; \; .
\end{equation}
It has to be
remarked that the coupling constant $g$ for the Thirring term 
shows up in the anomaly equation. This is due to the fact that 
it is the U(1) vector current which enters the Thirring term,
leading to an extra contribution to the anomaly.

As in QCD, the anomaly breaks the symmetry down to
$\mbox{SU(N)}_L \times \mbox{SU(N)}_R 
\times \mbox{U(1)}_V$. Since the right hand side
of the anomaly equation (11) is a divergence, the formal arguments 
\cite{bardeen} that
were applied in QCD to define a conserved current $\tilde{j}_5$ can be 
repeated. Thus when considering the symmetry properties, the
toy model is adequate for studying the problematic aspects in the
formulation of the U(1)-problem. 

%
%
\section{Outline of the construction}
In \cite{diss1} it was shown that for a perturbative treatment 
of the determinant of the massive Schwinger model
one would have to evaluate
infinitely many Feynman diagrams all of the same order in the 
fermion mass. Thus the
strategy will be to expand the mass term $\exp(-S_M[\overline{\psi},\psi])$ 
\begin{equation}
\langle P[\overline{\psi},\psi,A,h] \rangle = 
\frac{1}{Z}
\langle P[\overline{\psi},\psi,A,h] e^{-S_M[\overline{\psi},\psi]}
\rangle_0 = 
\frac{1}{Z} 
\sum_{n=0}^\infty \frac{(-1)^n}{n!}
\langle P[\overline{\psi},\psi,A,h] \; S_M[\overline{\psi},\psi]^n
\rangle_0 \; .
\end{equation}
The normalization constant $Z := \langle \exp \big( -S_M \big) \rangle_0$
also has to be expanded with respect to the fermion masses. It has to be
remarked that the Thirring term (6), as well as the cutoff $\Lambda$
are essential for this expansion.

The expectation values with subscript 0 showing
up in (12) are the expectation values of the massless model which are 
formally given by the path integral expression
\begin{equation}
\langle P[\overline{\psi},\psi,A,h] \rangle_0 \; := \;
\frac{1}{Z_0} \int {\cal D}h {\cal D}A {\cal D}\overline{\psi} {\cal D}\psi \; 
P[\overline{\psi},\psi,A,h] \; e^{-S_G[A] -S_h[h] 
-S_F [\overline{\psi},\psi,A,h]} \; .
\end{equation}
When integrating out the fermions (N flavors) one obtains 
$\mbox{det}[ \not\!\!{\partial} - 
i( e \not{\!\!\!A} + g^{1/2} \not{\!\!h}) ]^N$.
The fermion determinant 
is only defined when an ultraviolet and infrared
cutoff (for instance a finite space-time lattice, \cite{weingarten1, weingarten2})
is introduced. The determinant can then be normalized to 1 for $e, g = 0$, by
replacing it with $\mbox{det}[1 - K(A,h)]$ where
$K(A,h) = i \big(e\not{\!\!A} + g^{1/2}\not{\!h} \big)
{\not\!{\partial}}^{-1}$.
In two dimensions this determinant can be computed explicitey, 
using the idea of regularized fermion determinants 
(see e.g. \cite{seiler}). 
If we assume that the vector potentials $A_\mu$ and $h_\mu$ satisfy
some mild regularity and falloff conditions at infinity to make them
square integrable \cite{seiler}, the answer is
\begin{equation}
\mbox{det}[ 1 - K(A,h) ] \; = \; \exp \Big( -\frac{1}{2\pi} 
\Big\| eA^T \; + \; g^{1/2}h^T \Big\|_2^2 \Big) \; ,
\end{equation}
where $A_\mu^T$ and $h_\mu^T$ are the transverse parts of the vector fields
\begin{equation}
A_\mu^T = 
\Big( \delta_{\mu \nu} - \frac{\partial_\mu \partial_\nu}{\triangle} \Big)
A_\nu \; \; \; , \; \; \; h_\mu^T = 
\Big( \delta_{\mu \nu} - \frac{\partial_\mu \partial_\nu}{\triangle} \Big)
h_\nu \; .
\end{equation}
The action for $A_\mu$ as well as for $h_\mu$ are 
quadratic forms. Thus together with the logarithm of the
fermion determinant (14), the measure including the
effective action for the gauge field and 
the auxiliary field can be given a precise mathematical meaning in 
terms of Gaussian functional integrals.

Since Formula (14) for the fermion determinant mixes $h_\mu$
and $A_\mu$ we perform a shift of the auxiliary field   
\begin{equation}
h^\prime \; = \; 
h + \frac{e g^{1/2} N}{\pi + gN} T A \; ,
\end{equation}
in order to decouple the two fields. $T$ is the transverse 
projector defined in (7).
Thus we have 
\begin{equation}
\frac{1}{Z^0} {\cal D} h {\cal D}A 
e^{-S_G[A] -S_h[h]} \; 
\mbox{det}[ \not\!{\partial} - i( e \not{\!\!A} + g^{1/2} \not{\!h}) ]^N \; 
\; \leadsto \; \; d \mu_Q[A] d\mu_C[h^\prime] \; ,
\end{equation}
where $d \mu_Q[A]$ and $d\mu_C[h^\prime]$ denote Gaussian measures for
the gauge field $A_\mu$ and the shifted auxiliary field $h^\prime_\mu$, with 
covariances given by
\begin{equation}
Q _{\mu \nu} \; = \; \Big( -\triangle + \frac{e^2 N}{\pi + gN} \Big)^{-1} 
\; T_{\mu \nu} \; \; \; \; \; \; \; ,
\; \; \; \; \; \; \; 
C_{\mu \nu} \; = \; 
\frac{\pi}{\pi + gN} T_{\mu \nu} \; .
\end{equation}
Both expressions are quoted already in the transverse limit
i.e. $\lambda \rightarrow \infty$ and $\lambda^\prime \rightarrow \infty$ .

The final ingredient for the solution of massless $\mbox{QED}_2$ is the
fermion propagator $G(x,y;B)$ in an external field 
$B_\mu =  eA_\mu + g^{1/2}h_\mu$. The corresponding Green's function 
equation reads
\begin{equation}
\gamma_\mu \Big( \partial_\mu -iB_\mu (x) \Big) G(x,y;B) \; = \; 
\delta(x-y) \; .
\end{equation}
The solution was already found by Schwinger \cite{schwinger} and is given by
\begin{equation}
G(x,y;B) \; = \;  G^0(x-y) e^{i[\Phi(x) - \Phi(y)]}  \; ,
\end{equation}
where
\begin{equation}
\Phi(x) \; = \; -\int d^2z D(x-z) \Big( \partial_\mu B_\mu(z) + i \gamma_5
\varepsilon_{\mu \nu} \partial_\mu B_\nu(z) \Big) \; .
\end{equation}
$G^0(x)$ denotes the propagator for free massless fermions given
by $G^0(x) = \frac{1}{2\pi} \frac{\gamma_\mu x_\mu}{x^2}$, and
$\varepsilon_{\mu \nu}$ is the antisymmetric tensor. $D(x) = 
- \triangle^{-1}(x)$ denotes the propagator for massless bosons.
Expressing $h_\mu$ in terms of 
$h^\prime_\mu$ and using the fact that 
$A_\mu$ and $h^\prime_\mu$ are transverse fields 
one ends up with
\begin{equation}
G(x,y;A,h^\prime) \; = \;
\frac{1}{2\pi}\frac{1}{(x\!-\!y)^2}\!\left( \begin{array}{cc}
0 & e^{-[\chi(x)-\chi(y)]} \; 
\overline{(\tilde{x}\!-\!\tilde{y})} \\
e^{+[\chi(x)-\chi(y)]} \; (\tilde{x}\!-\!\tilde{y}) & 0
\end{array} \right)\; ,
\end{equation}
where
\begin{equation}
\chi(x) \; := \; 
\frac{\varepsilon_{\mu \nu} \partial_\mu}{\triangle} \left(
\frac{e \pi}{\pi + gN} A_\nu \; + \; g^{1/2} h^\prime_\nu \right)
\; \; \; \; ,
\; \; \; \; \tilde{x} := x_1 + ix_2 \; .
\end{equation}
As was discussed above the
construction of the determinant requires a falloff 
condition for the external fields $A_\mu, h_\mu$, which rules out 
nonvanishing winding number. They also require a UV-cutoff which we
impose as follows. The scalar field $\chi(x)$ at 
the single space-time point $x$ will be replaced by the convolute
\begin{equation}
\chi(x) = \; \longrightarrow \; 
\int \; d^2\xi \; \chi( \xi ) \; \delta_{n} ( \xi - x ) \; =: \;
\Big(\chi,\delta_n(x)\Big)\; ,
\end{equation}
where $\delta_n(x)$ denotes a $\delta$-sequence peaked at x
\begin{equation}
\delta_n(\xi-x) := 
\int \; \frac{d^2 p}{(2 \pi)^2} \; e^{ - \frac{|p|}{n} } \; e^{ip(\xi-x)} \; .
\end{equation}
Thus the propagator takes the form
\begin{equation}
G(x,y;A,h) = 
\frac{1}{2\pi}\frac{1}{(x-y)^2} \left( \begin{array}{cc}
0 & e^{-\big(\chi,\delta_n(x)-\delta_n(y)\big)} \; 
\overline{(\tilde{x}-\tilde{y})} \\
e^{+\big(\chi,\delta_n(x)-\delta_n(y)\big)} \; (\tilde{x}-\tilde{y}) & 0
\end{array} \right) \; .
\end{equation}
When one considers the limit $n \rightarrow \infty$ in the end, some
of the operators will have to be multiplied with a wave function
renormalization constant (see Eq. (42) below)
diverging as $n \rightarrow \infty$.

Since the dependence of the propagator on the external fields 
$A_\mu$ and $h_\mu^\prime$ is exponential, only Gaussian 
functional integrals are needed to solve the model.

%
%
\section{Construction of a unique vacuum}
It turns out that the expectation functional for the massless model
constructed so far violates clustering, and thus 
the vacuum state is not unique. In \cite{seilergatt} an explicit,
mathematically rigorous
definition of a new expectation functional $\langle .. \rangle_0^\theta$
which clusters, was given for $g=0$. We adapt this construction to  $g>0$
and compare the properties of $\langle .. \rangle_0^\theta$
with the expected properties of the $\theta$-vacuum of QCD.
%
%
\subsection{Violation of clustering}
To identify the operators that violate clustering, it is useful
(see \cite{seilergatt} for the $g=0$ case) to start with an
ansatz containing only the chiral densities 
$\overline{\psi}^{(b)} P_\pm \psi^{(b)} \; , \;
( P_\pm := (1 \pm \gamma_5)/2 )$
and discuss the effect of
adding vector currents and other modifications later.
Define
\begin{equation}
C(\tau) \; := \; C_1(\tau) \; - \; C_2  
\end{equation}
where
\[
C_1(\tau) \; := \;
\Big\langle \prod_{b=1}^N 
\prod_{i=1}^{n_b} \overline{\psi}^{(b)}(x^{(b)}_i\!+ \hat{\tau}) 
P_+ \psi^{(b)}(x^{(b)}_i\!+ \hat{\tau})
\prod_{i=1}^{m_b} \overline{\psi}^{(b)}(y^{(b)}_i\!+ \hat{\tau}) 
P_- \psi^{(b)}(y^{(b)}_i\! + \hat{\tau}) \]
\begin{equation} 
\times\;
\prod_{i=1}^{n^\prime_b} 
\overline{\psi}^{(b)}({x^\prime}^{(b)}_i) 
P_+ \psi^{(b)}({x^\prime}^{(b)}_i)
\prod_{i=1}^{m^\prime_b} 
\overline{\psi}^{(b)}({y^\prime}^{(b)}_i) 
P_- \psi^{(b)}({y^\prime}^{(b)}_i) \Big\rangle_0 
\end{equation}
and
\[ 
C_2 \; := \;
\Big\langle \prod_{b=1}^N 
\prod_{i=1}^{n_b} \overline{\psi}^{(b)}(x^{(b)}_i ) 
P_+ \psi^{(b)}(x^{(b)}_i )
\prod_{i=1}^{m_b} \overline{\psi}^{(b)}(y^{(b)}_i ) 
P_- \psi^{(b)}(y^{(b)}_i ) \Big\rangle_0 \]
\begin{equation}
\times \; \Big\langle \prod_{b=1}^N 
\prod_{i=1}^{n^\prime_b} 
\overline{\psi}^{(b)}({x^\prime}^{(b)}_i) 
P_+ \psi^{(b)}({x^\prime}^{(b)}_i)
\prod_{i=1}^{m^\prime_b} 
\overline{\psi}^{(b)}({y^\prime}^{(b)}_i) 
P_- \psi^{(b)}({y^\prime}^{(b)}_i) \Big\rangle_0 \; .
\end{equation}
$\hat{\tau}$ denotes the vector of length $\tau$ in 2-direction.
Violation of the cluster property now manifests itself in a nonvanishing 
limit
\begin{equation}
\lim_{\tau \rightarrow \infty} C(\tau) =: C  \neq 0 \; .
\end{equation}
It will be obtained for certain $n_b,m_b,n^\prime_b,m^\prime_b$.

$C_1(\tau)$ can be evaluated easily, since due to the exponential 
dependence of the fermion propagator on the external fields
it factorizes into the
expectation value for free massless fermions and an integral over the 
external fields      
\begin{equation}
C_1(\tau) \; = \; C_1(\tau)_{free} \times I(\tau) \; \; \; .
\end{equation}
Due to trace identities for the $\gamma$ algebra $C_1(\tau)_{free}$
does not vanish only for (compare \cite{seilergatt})
\begin{equation}
n_b + n^\prime_b = m_b + m^\prime_b \; \; \; , \; \; b = 1,...N \; .
\end{equation}
Using Cauchy's identity (see e.g. \cite{deutsch}) the result was 
computed in \cite{seilergatt}
\begin{equation}
C_1(\tau)_{free} = s
\left( \frac{1}{2\pi} \right)^{2\sum_b(n_n+n_b)}
\prod_{b=1}^N
\prod_{i,j = 1}^{n_b+n_b^\prime} 
\Big( w_i^{(b)} - z_j^{(b)} \Big)^{-2}
\prod_{1 \leq i < j \leq n_b + n_b^\prime}
\Big( w_i^{(b)} - w_j^{(b)} \Big)^2  
\Big( z_i^{(b)} - z_j^{(b)} \Big)^2 ,
\end{equation}
where $s$ denotes a sign depending on the $n^{(b)}, \; m^{(b)}$.
It is irrelevant for the following discussion.
The sets $\{w_j^{(b)}\},\{z_j^{(b)}\}$ for fixed flavor $b$ are 
given by
\[
\{ w_j^{(b)} \}_{j=1}^{n_b + n_b^\prime} := 
\{ x_l^{(b)} +\hat{\tau} , {x^\prime}^{(b)}_k \; | \; \;
l = 1,...n_b ; \; k = 1,...n^\prime_b \} \; ,
\]
\begin{equation}
\{ z_j^{(b)} \}_{j=1}^{m_b + m_b^\prime} :=
\{ y_l^{(b)} +\hat{\tau} , {y^\prime}^{(b)}_k \; | \; \;
l = 1,...m_b ; \; k = 1,...m^\prime_b \} \; .
\end{equation}
The integral over the external fields $A_\mu, h_\mu$ can be read off
from (26)
\[
I(\tau) = \int d\mu_{Q}[A] d\mu_{C}[h^\prime] \; 
\; \exp \Bigg(- 2\sum_{b=1}^N \Big[
\sum_{i=1}^{m_b} (\chi,\delta_n(x^{(b)}_i\!+ \hat{\tau}) +
\sum_{i=1}^{m^\prime_b} (\chi,\delta_n({x^\prime}^{(b)}_i) \Big] \Bigg) 
\]
\begin{equation}
\times \; \exp \Bigg(+ 2\sum_{b=1}^N \Big[
\sum_{i=1}^{n_b} (\chi,\delta_n(y^{(b)}_i\!+ \hat{\tau}) +
\sum_{i=1}^{n^\prime_b} (\chi,\delta_n({y^\prime}^{(b)}_i) \Big] \Bigg) \; .
\end{equation}
The result of inserting the $\delta$-sequence (25) 
and solving the Gaussian integrals 
can be written as 
\begin{equation}
I(\tau) \; = \; 
\exp \left( \sum_{i,j=1}^M V_n(w_i -  z_j) 
- \frac{1}{2}\sum_{i \neq j}^M V_n(w_i - w_j)
- \frac{1}{2}\sum_{i \neq j}^M V_n(z_i - z_j)  \right) \; .
\end{equation}
Again we introduced abbreviations for the involved 
space-time arguments given by 
\[
\{ w_j \}_{j=1}^M := 
\{ x_l^{(b)} +\hat{\tau} , {x^\prime}^{(b)}_k \; | \; \;
l = 1,...n_b ; \; k = 1,...n^\prime_b ; \; b = 1,...N \} \; ,
\]
\begin{equation}
\{ z_j \}_{j=1}^M :=
\{ y_l^{(b)} +\hat{\tau} , {y^\prime}^{(b)}_k \; | \; \; 
l = 1,...m_b ; \; k = 1,...m^\prime_b ; \; b = 1,...N \} \; .
\end{equation}
Due to (32) both sets contain the same number
$ M := \sum  (n_b + n_b^\prime) = \sum (m_b + m_b^\prime)$
of elements.
The potential $V_n$ showing up in (36) can easily be
obtained from the covariances
(18) and (25). It reads 
\[
V_n(x) \; = \; 4 \int \frac{d^2p}{(2\pi)^2} \; e^{-2\frac{|p|}{n}} \; 
\frac{e^2 \pi^2}{( \pi + g N )^2} \; \frac{1}{p^2 + e^2 N/(\pi + gN)}
\; \frac{1}{p^2} \; 
\Big( 1 - \cos(px) \Big) \]
\begin{equation}
+ \; 4 \int \frac{d^2p}{(2\pi)^2} \; e^{-2\frac{|p|}{n}} \; 
g \; \frac{\pi}{\pi + gN}
\frac{1}{p^2} \;
\Big( 1 - \cos(px) \Big) \; .
\end{equation}
In both integrals the infrared problem is cured by the 
$\Big( 1 - \cos(px) \Big)$ term. The first one even has no 
ultraviolet problem, and it can be solved after the limit 
$n \rightarrow \infty$ was taken. The other one has to be evaluated
for finite $n$. One obtains (see \cite{diss1} for the explicit 
computation of the integrals)
\[
V_n(x) = \frac{2}{\pi}
\frac{e^2 \pi^2}{( \pi\!+\!g N )^2}\frac{\pi\!+\!gN}{e^2 N}
\left( \ln|x| + \mbox{K}_0\left(\sqrt{\frac{e^2 N}{\pi\!+\!gN}} |x|\right)
+ \ln\left(\frac{1}{2} \sqrt{\frac{e^2 N}{\pi\!+\!gN}}\right)+\gamma\right)\]
\begin{equation}
 + \; \frac{2}{\pi} \; g \; \frac{\pi}{\pi + gN}\; 
\Bigg( \ln|x| + \ln\Big(\frac{n}{4}\Big) + O\Big(\frac{1}{n}\Big) \Bigg) 
\; = \; \frac{1}{N}\ln|x| + \tilde{V}(x) + 
\frac{2\pi g}{\pi\!+\!gN} \ln\left( \frac{n}{4} \right) + 
O\left(\frac{1}{n}\right) \; ,
\end{equation}
where we defined
\begin{equation}
\tilde{V}(x) := 
\frac{2\pi}{N(\pi + gN)} \;
\left(\mbox{K}_0\left(\sqrt{\frac{e^2 N}{\pi\!+\!gN}} |x|\right)
+ \ln\left(\frac{1}{2} 
\sqrt{\frac{e^2 N}{\pi\!+\!gN}}\right)+\gamma\right) \; .
\end{equation}
K$_0$ denotes the modified Bessel function and $\gamma$ is the 
Euler constant.
Thus one ends up with
\[ I(\tau) = \Big(\frac{n}{4}\Big)^{\frac{\pi g}{\pi + gN} \; 2 M} \; 
e^{O(\frac{1}{n})} 
\times
\exp \left( \sum_{i,j=1}^M \tilde{V}(w_i -  z_j) 
- \frac{1}{2}\sum_{i \neq j}^M \tilde{V}(w_i - w_j)
- \frac{1}{2}\sum_{i \neq j}^M \tilde{V}(z_i - z_j)  \right) \]
\begin{equation}
\times \prod_{i,j=1}^M \Big(w_i -  z_j\Big)^2 \; 
\prod_{i<j}^M \Big(w_i -  w_j\Big)^{-2} \Big(z_i -  z_j\Big)^{-2} \; .
\end{equation}
As announced, the factor diverging with $n$, 
the index of the $\delta$-sequence
(25) can be absorbed in a 
wave function renormalization constant $Z$ for the
chiral densities $\overline{\psi}^{(b)}P_\pm \psi^{(b)}$ 
\begin{equation}
Z \; := \left(\frac{4}{n}\right)^{\frac{\pi g}{\pi + gN}} \; .
\end{equation}
Putting together (33) and (41) one now can discuss the 
large-$\tau$ behaviour of $C_1(\tau)$. 
$\tilde{V}(x)$ showing up in the expression (41) 
for $I(\tau)$ depends on $x$ only via the modified Bessel function
$\mbox{K}_0$. Since $\mbox{K}_0$ approaches zero exponentially,
$\exp\big(\tilde{V}(x)\big)$ goes to a constant for large $\tau$, and the
only remaining $\tau$ dependence of $I(\tau)$ for $\tau \rightarrow \infty$
must come from the
rational function of the space-time arguments. Combining this with
$C_1(\tau)_{free}$ given in (33) one obtains
\begin{equation}
C_1(\tau) \; \; \propto \; \; \Big(\frac{1}{\tau^2}\Big)^E \;
\left( 1 + O\Big(\frac{1}{\tau}\Big) \right) \; ,
\end{equation}
where the exponent $E$ is given by
\begin{equation}
E \; = \; \sum_{a=1}^N (n_a\!-\!m_a)(n_a\!-\!m_a) -
\frac{1}{N} \sum_{a,b=1}^N
(n_a\!-\!m_a)(n_{b}\!-\!m_{b}) = 
\frac{1}{N}  \sum_{a,b=1}^N
(n_a\!-\!m_a)R_{ab}(n_{b}\!-\!m_{b}) \; .
\end{equation}
The matrix $R$ is defined as 
\begin{equation}
R_{ab} \; =: \; \delta_{ab} \; N \; - \; 1 \; .
\end{equation}
The corresponding eigenvalue problem can be solved easily.
One finds one eigenvalue 0, and $N-1$ eigenvalues $N$. The eigenvector
$x^0$ with the eigenvalue 0 is given by $x^0 = 1/\sqrt{N} (1,1,...1)^T$. 
Hence the quadratic form $x^T R x$ is positive semidefinite, and 
vanishes only if $x$ is a multiple of $x^0$. This implies that the
exponent $E$ is nonnegative and vanishes only for
\begin{equation}
n_b - m_b = m^\prime_b - n^\prime_b = n  \;\; \; \; \; \; 
\forall \; b \; = \; 1,2, \; ..... \; N \; , \; \;  \;\;  n \in 
\mbox{Z\hspace{-1.35mm}Z} \; .
\end{equation}
All those possibilities lead to a nonvanishing limit  
$C_1(\infty) := \lim_{\tau \rightarrow \infty} C_1(\tau)$.
In some of the cases $C_1(\infty)$ will be cancelled by $C_2$
which is given by (29). Using the trace identities for the
$\gamma$-algebra again (compare \cite{seilergatt}), one finds that 
$C_2$ does not vanish only for
\begin{equation}
n_b = m_b \; \; \; \; \mbox{and} \; \; \; \; 
n^\prime_b = m^\prime_b \; \; , \; \; \; \; b = 1,...N \; .
\end{equation}
In these cases $C_2$ then cancels $C_1(\infty)$ and the
operators cluster. Thus violation of clustering in $C(\tau)$
is expressed in the condition
\begin{equation}
n_b - m_b = m^\prime_b - n^\prime_b = n  \;\; , \;\;
\forall \; b = 1,2,..N \;\; , \;\;  n \in 
\mbox{Z\hspace{-1.35mm}Z} \setminus \{0\} \; .
\end{equation}
As was discussed in \cite{seilergatt} for the model without Thirring term,
the picture does not change when one inserts vector currents as well.
The only ingredient used for this result was the exponential 
dependence of the fermion propagator on the external fields, and the fact 
that the matrix $R$ defined in (45) is positive semidefinite. 
The same properties 
also hold for $g>0$ and the result of \cite{seilergatt} can be taken
over to the Schwinger-Thirring model.

In \cite{seilergatt} also the symmetry properties of the operators that 
violate clustering were discussed. Again the result can be taken over. 
Operators that violate clustering are singlets under 
$\mbox{U(1)}_V \times \mbox{SU(N)}_L\times \mbox{SU(N)}_R$, but
transform nontrivially under $\mbox{U(1)}_A$. 
%
%
\subsection{Definition of the clustering expectation functional 
$\langle .. \rangle_0^\theta$}
The symmetry properties discussed in the last section
can be used to decompose the vacuum functional into clustering states.
We adapt the prescription given in \cite{seilergatt}, involving
a limiting process which mimics the cluster procedure (30) to the case $g>0$.

Under a $\mbox{U(1)}_A$ transformation
\begin{equation}
\psi^{(b)} \; \longrightarrow \; e^{i \omega \gamma_5}
\psi^{(b)} \; ,
\end{equation}
an arbitrary monomial $\cal B$ of the fields transforms as 
\begin{equation}
{\cal B}({\{x\}}) \; \longrightarrow \;
e^{ i Q_5({\scriptstyle{\cal B}}) \omega} \; \; {\cal B}({\{x\}}) 
\; \; \; \; , \; \; \; \; Q_5({\scriptstyle{\cal B}}) \in
\mbox{Z\hspace{-.8ex}Z} \; .
\end{equation}
The new states $\langle .. \rangle^\theta_0$ labeled by a parameter 
$\theta \in [-\pi,\pi]$ are defined as follows
\begin{equation}
\langle \; {\cal B}({\{x\}}) \; \rangle^\theta_0 \; \; := \; \;
e^{i \theta \frac{Q_5({\scriptstyle{\cal B}})}{2N}} 
\lim_{\tau \rightarrow \infty}
\langle {\cal U}_\tau ({\scriptstyle {\cal B}}) \;  
{\cal B}({\scriptstyle\{x\}}) \rangle_0 \; .
\end{equation}
The new expectation value of an operator ${\cal B}$ is obtained by 
correlating ${\cal B}$ with a test operator ${\cal U_\tau}$ using the
old expectation functional and shifting ${\cal U_\tau}$ to timelike 
infinity. The result is multiplied with a phase which depends
on the chiral charge $Q_5({\scriptstyle{\cal B}})$.
The test operators ${\cal U}_\tau ({\scriptstyle {\cal B}})$ which also
depend on $Q_5({\scriptstyle{\cal B}})$
are defined as
\begin{equation}
{\cal U}_\tau ({\scriptstyle {\cal B}}) := 
\left\{ \begin{array}{l}
{\cal N}^{(n)} (\{y\})  \prod_{i=1}^n \prod_{b=1}^N
\overline{\psi}^{(b)}(y^{(b)}_i\!+\!\hat{\tau}) 
P_\mp\psi(y^{(b)}_i\!+\!\hat{\tau})
\; \; \; \; \mbox{for} \; 
\; \; Q_5({\scriptstyle{\cal B}}) = \pm 2nN , \; n \geq 1 \; , \\
\; \\
1 \; \; \mbox{otherwise} \; \; . \end{array} \right.
\end{equation}
Up to the requirement of being nondegenerate, 
the arguments $\{y^{(b)}_i\}$ are arbitrary. 
The normalizing factor ${\cal N}^{(n)} (\{y\})$ is defined such that 
\begin{equation} 
\lim_{\tau^\prime \rightarrow \infty} 
\langle {\cal U}_{\tau^\prime} ({\scriptstyle{\cal B}^\dagger})
\; {\cal U}_\tau ({\scriptstyle{\cal B}}) \rangle_0 \; = \; 1 \; .
\end{equation} 
It can be read off from (33), (41)
\[
{\cal N}^{(n)} (\{y\}) =  
\left( \frac{1}{2\pi} \right)^{-Nn} \!
\left[\frac{e^2 N}{4 ( \pi +gN)} e^{2\gamma} \right]^{
-\frac{\pi}{\pi + gN} \frac{Nn^2}{2}} \]
\begin{equation}
\times \;
\prod_{b=1}^N \prod_{1 \leq i < j \leq n} 
\left(y_i^{(b)}\!-\!y_j^{(b)}\right)^{-2}  
\exp \left( \frac{1}{2} \sum_{c,d=1}^N \sum_{k,l=1}^n 
(1-\delta_{cd} \delta_{kl})
\tilde{\tilde{V}}\left(y_k^{(c)} - y_l^{(d)} \right) \right) \; ,
\end{equation}
where
\begin{equation}
\tilde{\tilde{V}}(x) \; := \; \frac{1}{N}\ln(x^2) + \tilde{V}(x) \; .
\end{equation}
In \cite{seilergatt} the following theorem was shown to hold
\vskip5mm
\noindent
{\bf Theorem 1 :}\\
{\bf i) } The cluster decomposition property holds for 
$\langle \; .. \; \rangle^\theta_0$ .\\
{\bf ii) } The state $\langle .. \rangle_0$ constructed initially
is recovered by averaging over $\theta$
\begin{equation}
\langle \; .. \; \rangle = \frac{1}{2\pi}\int_{-\pi}^{\pi}
\langle \; .. \; \rangle^\theta_0 \; d\theta \;.
\end{equation}
\vskip5mm
\noindent
The theorem establishes that our prescription (51) indeed ensures 
clustering and thus leads to a unique vacuum. 
Thus $\langle .. \rangle^\theta_0$ is exactly what is hoped to
have been obtained by the expression (1) 
for the $\theta$-vacuum of QCD.
The prescription (51) is mathematically rigorous, whereas
(1) has a more formal character. Nevertheless it is interesting to
compare the properties of the two constructs.
This opens a series of lessons for the topics discussed in
the introduction that we will draw from the model.
\vskip5mm
\noindent
{\bf Lesson 1 :} (On the vacuum structure of QED$_2$.)
\vskip3mm
\noindent 
{\it The newly defined vacuum functional $\langle .. \rangle_0^\theta$
gives rise to the chirality selection rule}
\begin{equation}
\langle {\cal B} \rangle^\theta_0 \; = \; 0 \; \; \; \mbox{unless} \; \; \; 
Q_5({\scriptstyle{\cal B}}) \; = \; 2 N \; n \;
\; \; , \; n \; \in \; \mbox{Z\hspace{-1.35mm}Z} \; .
\end{equation}
In particular operators ${\cal B}$ with nonvanishing
$\langle {\cal B} \rangle^\theta_0$ have to be singlets under
$\mbox{U(1)}_V \times \mbox{SU(N)}_L\times \mbox{SU(N)}_R$.
The property (57) can be seen to hold from the discussion of the behaviour of 
$C_1(\tau)$ in Section 4.1. Eq. (46) immediately leads to (57).

Thus the chiral condensate $\langle \overline{\psi} \psi
\rangle$ for N=1 generalizes in the case of 
several flavors to e.g. 
\begin{equation}
\langle \; \prod_{b=1}^N \overline{\psi}^{(b)} P_\pm \psi^{(b)}
\rangle_0^{\theta} \; \neq \; 0 \; .
\end{equation}
This can be understood easily using Coleman's theorem
\cite{colemangold}, allowing no spontaneous breaking of
SU(N)$_A$ in two dimensions (only U(1)$_A$ is explicitly broken
by the anomaly). Thus the expectation values of
all operators that do not transform trivially under SU(N)$_A$
(such as $\langle \overline{\psi}^{(b)} \psi^{(b)}\rangle$
for some flavor $b$) have to vanish. The operator in (58) 
does not fall into this class and acquires a nonvanishing expectation 
value. Chiral condensates of the type
(58) were also discussed in \cite{rotheswi}, \cite{belvedere},
where they were 
related to classical gauge field configurations with winding number. 
This is an approach which we avoid here.
It has been demonstrated (see \cite{joos1}-\cite{wipf6},
\cite{smilga} for some recent work) 
that putting the model on a finite torus 
leads to breaking of the chiral symmetry and to a nonvanishing
condensate $\langle \overline{\psi}^{(b)} \psi^{(b)}\rangle$.

A selection rule equivalent to (57) can formally be obtained for
QCD (see e.g. the review article \cite{peccei}) from the 
naive construction \cite{callan1, callan2} of the $\theta$-vacuum of QCD.
Anyway this result is too naive, since the 
chiral symmetry is believed to be broken spontaneously in QCD
(There is only one instance of a model relevant for QCD, 
where chiral symmetry breaking is proven \cite{salm}.). 
In fact
$\langle \theta | \overline{u} u | \theta \rangle \neq 0$
which is the manifestation of chiral symmetry breaking,
is one of the main assumptions of current algebra
(see \cite{crewtherkai} for a review).

%
%
\section{Bosonization and the GSG}
In this section it will be shown
that expectation values of chiral densities and certain
currents within the $\langle \; .. \; \rangle^\theta_0$ vacuum functional 
can be bosonized. For the massive model this gives rise to
a generalized Sine-Gordon model (GSG).

\subsection{Evaluation of a generating functional}
In order to establish the bosonization we evaluate a generating functional 
where the operators 
$\overline{\psi}^{(b)} P_\pm \psi^{(b)} \; , \; j^{(b)}_\mu$ 
which we want to bosonize enter. It reads
\begin{equation}
E(n_b,m_b;a^{(b)}) := 
\Big\langle
\prod_{b=1}^N
\prod_{i=1}^{n_b} 
\overline{\psi}^{(b)}({x}^{(b)}_i) 
P_+ \psi^{(b)}({x}^{(b)}_i)
\prod_{j=1}^{m_b} 
\overline{\psi}^{(b)}({y}^{(b)}_j) 
P_- \psi^{(b)}({y}^{(b)}_j) 
e^{ie\sum_{c=1}^N (a_\mu^{(c)},j_\mu^{(c)})} \Big\rangle_0^\theta \; .
\end{equation}
Obviously this is a simple modification of expectation values already 
considered in the last section. Only a generating exponential has been added
where
the vector currents $j_\mu^{(b)}$ for the various flavors $b$,
couple to external sources $a_\mu^{(b)}$. Since those 
sources couple in the same way as the gauge and the auxiliary field, 
they can be included into the fermion determinant 
giving rise to (see (14))
\begin{equation}
\prod_{b=1}^N \; \exp\Big( - \frac{1}{2\pi} \Big\|
eA^T \; + \; g^{1/2}h^T \; + \; e a^{(b)} \Big\|^2_2 \Big) \; .
\end{equation}
Thus we simply obtain an extra factor in the functional integral over
$A_\mu$ and $h_\mu$ (respectively $h^\prime_\mu$).

To work out the dependence on the
sources $a^{(b)}_\mu$ we first consider the case
\begin{equation}
n_b - m_b = 0 \; \; \; , \; \; \; \; b = 1,2, \; ... \; , N \; ,
\end{equation}
where the evaluation of $\langle .. \rangle^\theta_0$
is remarkably simple, i.e. it coincides
with the naive expectation value (compare (51)). Again the expression 
factorizes 
\begin{equation}
E(n_b,n_b;a^{(b)}) := I(n_b,n_b;a^{(b)})
\times E_{free} (n_b,n_b) \; ,
\end{equation}
and the factor from the functional integral reads
\[ 
I(n_b,n_b;a^{(b)}) = \int d\mu_Q[A] d\mu_C[h^\prime] \; \; 
\exp \left( - 2 \sum_{b=1}^N \sum_{j=1}^{n_b}
\Big( \chi^{(b)}, \delta_n(x_j^{(b)})\!-\!\delta_n(y_j^{(b)}) \Big) \right)
\]
\begin{equation}
\times \;
\exp\left( -\frac{e^2}{\pi\!+\!gN} \Big( A, T \sum_{b=1}^N a^{(b)} \Big)
-\frac{e \sqrt{g}}{\pi} \Big( h^\prime , T \sum_{b=1}^N a^{(b)} \Big)
-\frac{e^2}{2\pi} \sum_{b=1}^N \Big( a^{(b)} , T a^{(b)} \Big) \right) \; .
\end{equation}
The first exponent consists of a sum over the inner products of the 
$\delta$-sequences centered at the various space-time arguments with
the gauge and the auxiliary fields which enter $\chi^{(b)}$.
The extra flavor superscript in $\chi^{(b)}$ is due to the
fact, that the external sources $a^{(b)}_\mu$ were included into
the fermion action and thus show up also in the propagator.
Hence the definition (23) for $\chi$ has to be generalized to
\begin{equation}
\chi^{(b)} \; := \; 
\frac{\varepsilon_{\mu \nu} \partial_\mu}{\triangle} \left(
\frac{e \pi}{\pi + gN} A_\nu \; + \; g^{1/2} h^\prime_\nu \right) \; 
+ \; e \frac{\varepsilon_{\mu \nu} \partial_\mu}{\triangle}a^{(b)}_\nu \; .
\end{equation}
In the propagator (26), $\chi$ has to be replaced by $\chi^{(b)}$
which in turn immediately leads to the first exponent in (63).
The exponent in the second line of (63) stems from the determinant (60)
and contains only a term linear in $A_\mu$ and $h^\prime_\mu$, 
since the quadratic 
term is already included in the Gaussian measure.

\noindent
The functional integral (63) can now be evaluated. The result is an 
exponential which mixes the sources and the $\delta$-sequences.
If the terms quadratic in the sources $a^{(b)}$ are collected,
one finds that the corresponding term can be written as
\begin{equation}
-\frac{e^2}{2\pi} \sum_{b,c=1}^N 
\Big(\varepsilon_{\mu \nu} \partial_\mu a^{(b)}_\nu, M_{bc} \;
\varepsilon_{\rho \sigma} \partial_\rho a^{(c^\prime)}_\sigma \Big) \; ,
\end{equation}
where the covariance $M$ is given by
\begin{equation}
M = \frac{1}{-\triangle  + e^2N/(\pi + gN)} 
\left[ \frac{\pi}{\pi + gN} \mbox{1\hspace{-.6ex}I} 
+ \frac{g}{\pi + gN} R \right]
\; + \; \frac{e^2}{\pi + gN} \frac{-1}{\triangle}
\frac{1}{-\triangle  + e^2N/(\pi + gN)} R \; .
\end{equation}
The numerical matrix $R$ was introduced in (45).
As already discussed the eigenvalues of $R$ are 0, and $N$, where 
the latter is $N\!-\!1$-fold degenerate. Diagonalization of $R$ thus 
leads to
\begin{equation}
U R U^T \; = \; \mbox{diag} ( 0 , N , N , \; .... \;N ) \; ,
\end{equation}
where the orthogonal matrix $U^T$ is given by 
\begin{equation}
U^T := \left( 
c^{(1)} \! \left[ \begin{array}{c} 
1 \\ 1 \\ . \\ .\\  . \\ . \\ 1 \end{array}  \right] ,
c^{(2)} \! \left[ \begin{array}{c} 
1 \\ . \\ . \\ .\\  . \\ 1 \\ -(N\!-\!1) \end{array}   \right] ,
c^{(3)} \! \left[ \begin{array}{c} 
1 \\ . \\ . \\ .\\  1 \\ -(N\!-\!2) \\ 0 \end{array}   \right]
\; . \; . \; . \; 
c^{(N)} \! \left[ \begin{array}{c} 
\; 1 \\ -1 \\ 0 \\ .\\  . \\ 0 \\ 0 \end{array}   \right]
\right)  .
\end{equation}
The normalization constants $c^{(I)}$ are given by
\[
c^{(1)} := \frac{1}{\sqrt{N}} \; \; , \; \;
c^{(2)} := \frac{1}{\sqrt{N-1+(N-1)^2}} \; , \]
\begin{equation}
c^{(3)} := \frac{1}{\sqrt{N-2+(N-2)^2}} \; \; , 
\; \; . \; .\; . \; , \; . \; .\; . \; , \; \; 
c^{(N)} := \frac{1}{\sqrt{2}} \; .
\end{equation}
(67) allows to express the quadratic form (66) in terms of a covariance
$K$ diagonal in flavor
\begin{equation}
K \; := \; U \; M \; U^T \; \; \Longleftrightarrow \; \;
M \; = \; U^T \; K \; U \; .
\end{equation}
Explicitly $K$ is given by
\begin{equation}
K = \left( \begin{array}{ccccc}
\frac{\pi}{\pi\!+\!gN} \frac{1}{-\triangle\!+\!e^2N/(\pi\!+\!gN)} 
& & & & \\
 & \frac{1}{-\triangle} & & & \\
 & & . & & \\
 & & & . & \\
 & & & & \frac{1}{-\triangle} \end{array} \right) \; .
\end{equation}
Obviously the covariance $K$ describes one
massive and $\mbox{N}-1$ massless particles. 
Finally the term quadratic in the sources reads
\begin{equation}
-\frac{e^2}{2\pi} 
\Big( \varepsilon_{\mu \nu} \partial_\mu U a_\nu, K \;
\varepsilon_{\rho \sigma} \partial_\rho U a_\sigma \Big) \; ,
\end{equation}
where matrix notation in flavor space was used. 
Eq. (72) suggests to define new sources $A_\mu^{(I)}$ that are linear
combinations of the $a_\mu^{(b)}$
\begin{equation}
A_\mu^{(I)} := \sum_{b=1}^N U_{I b} a_\mu^{(b)} \; \; 
\stackrel{U^T = U^{-1}}{\Longleftarrow \!\!=\!\!=\!\!=\!\!=\!\!= 
\!\!\! \Longrightarrow}  \; \;
a_\mu^{(b)} = \sum_{I=1}^N U_{I b} A_\mu^{(I)} \; .
\end{equation}
Rewriting the coupling term in $E(n_b,n_b;a^{(b)})$
allows the identification of the
currents $J_\mu^{(I)}$ that couple to the new sources $A_\mu^{(I)}$
\begin{equation}
\sum_{b=1}^N \Big( a_\mu^{(b)},j_\mu^{(b)} \Big) \; = \;
\sum_{I=1}^N \sum_{b=1}^N 
\Big( U_{I b} A_\mu^{(I)}, j_\mu^{(b)} \Big) \; := \;
\sum_{I=1}^N \Big( A_\mu^{(I)}, J_\mu^{(I)} \Big) \; ,
\end{equation}
where we defined
\begin{equation}
J_\mu^{(I)} \; := \; \sum_{b=1}^N U_{I b} j_\mu^{(b)} \; \; \;
\; \; \; I = 1,2, ... N \;.
\end{equation}
Inspecting the explicit form of the matrix $U$ defined in (68)
one can express the new currents also as
\begin{equation}
J_\mu^{(I)} := \sum_{b,c=1}^N \;
\overline{\psi}^{(b)} \gamma_\mu H^{(I)}_{bc} \; \psi^{(c)}
\end{equation}
where the $\mbox{N}\times \mbox{N}$ matrices $H^{(I)}$ 
are generators of a Cartan subalgebra of $\mbox{U(N)}_{flavor}$, and thus 
the currents (75) will be referred to as Cartan currents in the following.
Using (74) one can define a new generating functional $E(n_b,m_b;A^{(I)})$
which now contains the 
currents $J^{(I)}_\mu$ coupled to the sources $A^{(I)}_\mu$
\begin{equation}
E(n_b,m_b;A^{(I)}) := 
\Big\langle
\prod_{b=1}^N
\prod_{i=1}^{n_b} 
\overline{\psi}^{(b)}({x}^{(b)}_i) 
P_+ \psi^{(b)}({x}^{(b)}_i)
\prod_{j=1}^{m_b} 
\overline{\psi}^{(b)}({y}^{(b)}_j) 
P_- \psi^{(b)}({y}^{(b)}_j) 
e^{ie\sum_{I=1}^N (A_\mu^{(I)},J_\mu^{(I)})} \Big\rangle_0^\theta \; .
\end{equation}
Using (65) and (33) for $E_{free} (n_b,n_b)$ showing up in 
(62) one obtains (still $n_b = m_b$)
\[
E(n_b,n_b;A^{(I)}) = 
\Big( \frac{1}{2\pi} \Big)^{2 \sum_b n_b}
\times \exp \left( -\frac{e^2}{2\pi} \sum_{I=1}^N 
\Big( \varepsilon_{\mu \nu} \partial_\mu A_\nu^{(I)} , K_{I I}
\varepsilon_{\rho \sigma} \partial_\rho A^{(I)}_\sigma \Big) \right)
\]
\begin{equation}
\times \prod_{b=1}^N \prod_{j=1}^{n_b} \exp \left(
2e \sum_{I=1}^N \left( \varepsilon_{\mu \nu} \partial_\mu A^{(I)}_\nu,
K_{I I} U_{I b} 
\Big[ \delta_n(x_j^{(b)}) - \delta_n(y_j^{(b)}) \Big] \right) \right) 
\times \rho_n ( \{x^{(b)}_j \}, \{y^{(b)}_j \} ) \; .  
\end{equation}
$\rho_n( \{x^{(b)}_j \}, \{y^{(b)}_j \} )$ denotes the 
factor that depends on the space-time arguments. Furthermore
it still depends on $n$, the index of the $\delta$-sequences (25).
The wave function renormalization (42) for the chiral densities
has to be applied before the limit $n \rightarrow \infty$ is
taken. We will quote the explicit form of $\rho_\infty$ later.

The result (78) now can be generalized to 
\begin{equation}
n_b - m_b = l \; \; \; \; \; \; \; \; \;
l \in \mbox{Z\hspace{-.8ex}Z}
\; \; \; \; \; \; \; \; \; b = 1,2, \; ... \; , N \; ,
\end{equation}
which covers all cases where the expectation functional
$\langle .. \rangle_0^\theta$ gives nonvanishing results for 
$E(n_b,m_b; A^{(I)})$.
The result can be obtained easily by following the 
argumentation given in the last section. In fact the term quadratic in 
the sources is not affected by the $\theta$-prescription, and 
$\rho_n( \{x^{(b)}_j \}, \{y^{(b)}_j \} )$ can be read off from
(33), (41) immediately. Only the term that mixes the sources with the
space time arguments of the chiral densities has to be generalized,
but this is straightforward. One ends up with
\[
E(n_b,m_b;A^{(b)}) 
\; 
= \; \exp \left( -\frac{e^2}{2\pi} \sum_{I=1}^N 
\Big( \varepsilon_{\mu \nu} \partial_\mu A_\nu^{(I)} , K_{I I} \;
\varepsilon_{\rho \sigma} \partial_\rho A^{(I)}_\sigma \Big) \right) 
\]
\[
\times \prod_{b=1}^N 
\prod_{j=1}^{n_b} \exp \left(
2e \sum_{I=1}^N \Big( \varepsilon_{\mu \nu} \partial_\mu A^{(I)}_\nu,
K_{I I} \; U_{I b} \; \delta(x_j^{(b)}) \Big) \right) \]
\begin{equation}
\times \prod_{b=1}^N 
\prod_{j=1}^{m_b} \exp \left(
- 2e \sum_{I=1}^N \Big( \varepsilon_{\mu \nu} \partial_\mu A^{(I)}_\nu,
K_{I I} \; U_{I b} \; \delta(y_j^{(b)}) \Big) \right)
\times \rho_\infty ( \{x^{(b)}_j \}, \{y^{(b)}_j \} ) \; ,  
\end{equation}
where $\rho_\infty ( \{x^{(b)}_j \}, \{y^{(b)}_j \} )$ is given by 
(the limit $n \rightarrow \infty$ and the wave function renormalization
(42) have already been performed)
\[
\rho_\infty ( \{x^{(b)}_j \}, \{y^{(b)}_j \} ) \; = \]
\[
h(n_b,m_b) \; 
\Big(\frac{1}{2\pi}\Big)^{\sum_b (n_b + m_b)} \; \;
e^{ i \frac{\theta}{N} \sum_b (n_b - m_b)} 
\;
\times \; \left( e \sqrt{\frac{N}{\pi\!+\!gN}} \frac{e^\gamma}{2} 
\right)^{\frac{\pi}{\pi + gN} \frac{1}{N} 
\Big( \sum_b (n_b - m_b) \Big)^2} \]
\[
\times \exp \left( 
\sum_{b, c=1}^N \sum_{j=1}^{n_b} \sum_{l=1}^{m_{b^\prime}}
\tilde{\tilde{V}}(x^{(b)}_j - y^{(c)}_{l}) \right) \]
\[
\times \exp \left( -\frac{1}{2}
\sum_{b,c=1}^N \left(
\sum_{j=1}^{n_b} \sum_{l}^{n_{c}}
\Big(1\!-\!\delta_{b c} \delta_{j l} \Big)
\tilde{\tilde{V}}(x^{(b)}_j - x^{(c)}_{l}) 
\; + \; \sum_{j=1}^{m_b} \sum_{l=1}^{m_{c}}
\Big(1\!-\!\delta_{bc} \delta_{jl} \Big)
\tilde{\tilde{V}}(y^{(b)}_j - y^{(c)}_{l}) \right) \right)\]
\[
\times \exp \left( 
- \sum_{b=1}^N \sum_{j=1}^{n_b} \sum_{l=1}^{m_{b}}
\ln (x^{(b)}_j - y^{(b)}_{l})^2 \right) \]  
\begin{equation}
\times \exp \left( \frac{1}{2}
\sum_{b=1}^N \left(
\sum_{j=1}^{n_b} \sum_{l=1}^{n_{b}}(1\!-\!\delta_{j l})
\ln (x^{(b)}_j - x^{(b)}_{l})^2 
\; + \;
\sum_{j=1}^{m_b} \sum_{l=1}^{m_{b}}(1\!-\!\delta_{j l})
\ln (y^{(b)}_j - y^{(b)}_{l})^2 \right) \right)\; .
\end{equation}
The factor $h(n_b,m_b)$ is defined as
\begin{equation}
h(n_b,m_b) \; := \;
\sum_{l= -\infty}^{+\infty} \prod_{b=1}^N \delta_{n_b\!-\!m_b, l} \; \; \; \; .
\end{equation}
It is equal to one whenever 
the prescription (51) allows a nonvanishing result for
$E(n_b,m_b;A^{(b)})$, otherwise it is zero. $\tilde{\tilde{V}}$ is given by
(55).
Expression (80) can now be used to identify the correct bosonization.
%
%
\subsection{Bosonization prescription}
Bosonization means that the generating functional 
$E(n_b,m_b;A^{(b)})$ can also be obtained by computing the
vacuum expectation value of a certain functional 
${\cal F}(n_b,m_b;A^{(b)};\Phi^{(I)})$ depending on fields $\Phi^{(I)}$
in a bosonic theory. Every operator that was used to 
define $E(n_b,m_b;A^{(b)})$ 
will have a transcription
in terms of the $\Phi^{(I)}$ which then enters 
${\cal F}(n_b,m_b;A^{(b)};\Phi^{(I)})$.
Inspecting (80) makes it plausible to try it with Gaussian fields
$\Phi^{(I)}$ with some covariances $Q^{(I)}$ which are related to the
$K_{I I}$ (see (71)).
Thus one can express the idea of bosonization in the following
formula
\begin{equation}
E(n_b,m_b;A^{(b)}) \; = \;
\Big\langle {\cal F} (n_b,m_b;A^{(b)};\Phi^{(I)}) 
\Big\rangle_{ \{Q^{(I)}\} } \; ,
\end{equation}
where $\langle .. \rangle_{ \{Q^{(I)} \} }$ denotes expectation values
for the fields $\Phi^{(I)}$ with respect to the covariances $Q^{(I)}$.
Two steps have to be done. First define an appropriate covariance
$Q^{(I)}$ and then establish the correct transcription of the fermionic 
operators into bosonic ones.

The definition of the $Q^{(I)}$ is rather simple. We define
\begin{equation} 
Q^{(1)} \; \; := \; \; \frac{1}{-\triangle + e^2 N/(\pi + gN)} \; \; = \; \;  
\frac{\pi + gN}{\pi} K_{1 1} \; ,
\end{equation}
and
\begin{equation}
Q^{(I)} \; := \; \frac{1}{-\triangle} \; = \; K_{I I} \; \;
\; \;  \; I = 2, ...N \; .
\end{equation}
Thus the $Q^{(I)}$ are just the canonically normalized $K_{I I}$.
The term in (80) which is quadratic in the sources $A^{(b)}$ then implies
the following prescription for the bosonization of the Cartan 
currents (see also \cite{belvedere} for the $g=0$ case)
\begin{equation}
J^{(I)}_\nu(x) \; \longleftrightarrow \; 
\left\{ \begin{array}{ll}  
- \frac{1}{\sqrt{\pi\!+\!gN}} \; 
\varepsilon_{\mu \nu} \partial_\mu \Phi^{(1)}(x) \; \; \; \; \; \; & I = 1 \\
\; & \;   \\
- \frac{1}{\sqrt{\pi}} \;
\varepsilon_{\mu \nu} \partial_\mu \Phi^{(I)}(x) \; \; \; \; \; \;
& I = 2,\; ... \; N \; .
\end{array} \right.
\end{equation}
With this choice the term linear in $A^{(b)}_\mu$ already fixes the
structure of the transcription of the chiral densities to
(see also \cite{hosotani1, hosotani2})
\[
\overline{\psi}^{(b)}(x) P_\pm \psi^{(b)}(x) \; \longleftrightarrow \]
\begin{equation}
\frac{1}{2\pi} c^{(b)} 
: e^{\mp i 2 \sqrt{\pi} \sqrt{\frac{\pi}{\pi +gN}} U_{1 b} \Phi^{(1)}(x)} 
:_{M^{(1)}} \; \prod_{I=2}^N 
: e^{\mp i 2 \sqrt{\pi} U_{I b} \Phi^{(I)}(x)} 
:_{M^{(I)}} e^{\pm i \frac{\theta}{N}}\; ,
\end{equation}
as can be seen from the exponents in (80) linear in the
sources. Here 
$: .. :_{M^{(I)}}$ denotes normal ordering with respect to mass
$M^{(I)}$ (see e.g. \cite{simonphi}). Those normal ordering masses as well 
as the real numbers $c^{(b)}$ are free parameters that will be fixed
later.
Inserting the prescriptions (86), (87) into the definition of 
$E(n_b,m_b;A^{(b)})$ one obtains
\[
E(n_b,m_b;A^{(b)}) \; \longleftrightarrow \; 
\left(\frac{1}{2\pi}\right)^{\sum_b (n_b + m_b)} \; \;
e^{ i \frac{\theta}{N} \sum_b (n_b - m_b)} \; \;
\prod_{b=1}^N \Big( c{(b)} \Big)^{n_b + m_b} 
\]
\[
\Bigg\langle \; \; \prod_{b=1}^N \prod_{j=1}^{n_b} \left[
: e^{- i 2 \sqrt{\pi} \sqrt{\frac{\pi}{\pi +gN}} U_{1 b} 
\Phi^{(1)}(x^{(b)}_j)} :_{M^{(1)}} \; \prod_{I=2}^N 
: e^{- i 2 \sqrt{\pi} U_{I b} \Phi^{(I)}(x^{(b)}_j)} 
:_{M^{(I)}} \right] 
\]
\[
\times \; \prod_{b=1}^N \prod_{j=1}^{m_b} \left[
: e^{+ i 2 \sqrt{\pi} \sqrt{\frac{\pi}{\pi +gN}} U_{1 b} 
\Phi^{(1)}(y^{(b)}_j)} :_{M^{(1)}} \; \prod_{I=2}^N 
: e^{+ i 2 \sqrt{\pi} U_{I b} \Phi^{(I)}(y^{(b)}_j)} 
:_{M^{(I)}} \right]
\]
\begin{equation}
\times \; \exp \left( - \frac{i e}{\sqrt{\pi + gN}} 
\Big( A_\mu^{(1)}, \varepsilon_{\nu \mu} \partial_\nu \Phi^{(1)} \Big)
- \frac{i e}{\sqrt{\pi}} \sum_{I=2}^N
\Big( A_\mu^{(I)}, \varepsilon_{\nu \mu} \partial_\nu \Phi^{(I)} \Big)
\right) \; \; \Bigg\rangle_{ \{ Q^{(I)} \} } \; .
\end{equation}
The Gaussian integrals can be solved easily since they factorize
with respect to the $\Phi^{(I)}$. One then obtains for the right hand
side of the last equation
\[
\exp \left( - \frac{e^2}{2 \pi} \Big(
\varepsilon_{\mu \nu} \partial_\mu A_\nu^{(1)}, 
\frac{\pi}{\pi\!+\!gN} Q^{(1)}
\varepsilon_{\rho \sigma} \partial_\rho A_\sigma^{(1)} \Big)
- \frac{e^2}{2 \pi} \sum_{I=2}^N \Big(
\varepsilon_{\mu \nu} \partial_\mu A_\nu^{(I)}, Q^{(I)}
\varepsilon_{\rho \sigma} \partial_\rho A_\sigma^{(I)} \Big) \right) 
\]
\[
\times \prod_{b=1}^N \prod_{j=1}^{n_b} \exp \left( 
+ 2e U_{1b} \Big( \varepsilon_{\mu \nu} \partial_\mu A_\nu^{(1)},
\frac{\pi}{\pi+gN} Q^{(1)} \delta(x^{(b)}_j) \Big) \right)
\]
\[
\times \prod_{b=1}^N \prod_{j=1}^{n_b} \exp \left( 
+ 2e \sum_{I=2}^N U_{Ib} \Big( \varepsilon_{\mu \nu} \partial_\mu A_\nu^{(I)},
Q^{(I)} \delta(x^{(b)}_j) \Big) \right)
\]
\[
\times \prod_{b=1}^N \prod_{j=1}^{m_b} \exp \left( 
- 2e U_{1b} \Big( \varepsilon_{\mu \nu} \partial_\mu A_\nu^{(1)},
\frac{\pi}{\pi+gN} Q^{(1)} \delta(y^{(b)}_j) \Big) \right)
\]
\begin{equation}
\times \prod_{b=1}^N \prod_{j=1}^{m_b} \exp \left(
- 2e \sum_{I=2}^N U_{Ib} 
\Big( \varepsilon_{\mu \nu} \partial_\mu A_\nu^{(I)},
Q^{(I)} \delta(y^{(b)}_j) \Big) \right)
\; \times \; \rho_B(\{x_j^{(b)}\}, \{y_j^{(b)}\}) \; .
\end{equation}
Comparing (80) and (89) shows immediately that the terms
quadratic and linear in the sources $A^{(I)}$ come out correctly. Thus 
there is left to show
\begin{equation}
\rho_B(\{x_j^{(b)}\}, \{y_j^{(b)}\}) \; \; = \; \; 
\rho_\infty(\{x_j^{(b)}\}, \{y_j^{(b)}\}) \; ,
\end{equation}
where $\rho_\infty(\{x_j^{(b)}\}, \{y_j^{(b)}\})$ is given by (81).
As mentioned before, the integral over the $\Phi^{(I)}$ factorizes
such that $\rho_B(\{x_j^{(b)}\}, \{y_j^{(b)}\})$ reads
\[
\rho_B(\{x_j^{(b)}\}, \{y_j^{(b)}\}) = 
\left(\frac{1}{2\pi}\right)^{\sum_b (n_b + m_b)} \; \;
e^{ i \frac{\theta}{N} \sum_b (n_b - m_b)} \; \;
\prod_{b=1}^N \Big( c^{(b)} \Big)^{n_b + m_b} 
\]
\[
\times \Bigg\langle
\prod_{b=1}^N \; \prod_{j=1}^{n_b} 
: e^{- i 2 \sqrt{\pi} \sqrt{\frac{\pi}{\pi +gN}} U_{1 b} 
\Phi^{(1)}(x^{(b)}_j)} :_{M^{(1)}} \; 
\prod_{l=1}^{m_b} 
: e^{+ i 2 \sqrt{\pi} \sqrt{\frac{\pi}{\pi +gN}} U_{1 b} 
\Phi^{(1)}(y^{(b)}_l)} :_{M^{(1)}} \Bigg\rangle_{Q^{(1)}} 
\]
\begin{equation}
\times \prod_{I=2}^N \;
\Bigg\langle
\prod_{b=1}^N \; \prod_{j=1}^{n_b} 
: e^{- i 2 \sqrt{\pi} U_{I b} 
\Phi^{(I)}(x^{(b)}_j)} :_{M^{(I)}} \; 
\prod_{l=1}^{m_b} 
: e^{+ i 2 \sqrt{\pi} U_{I b} 
\Phi^{(I)}(y^{(b)}_l)} :_{M^{(I)}} \Bigg\rangle_{Q^{(I)}}
\end{equation}
The expectation values of the normal ordered exponentials are 
rather simple to evaluate. One only has to take
care of the neutrality condition (see e.g. \cite{frohlich1})
which has to be fulfilled for 
normal ordered exponentials of a massless scalar field $\Phi$
\begin{equation}
\lim_{\mu \rightarrow 0} \; 
\langle \prod_{j=1}^{n} : e^{i \Phi(t_j)} :_M  \rangle_{C^\mu}
\; = \; \left\{ \begin{array}{cc}
e^{+\frac{1}{2} \sum_{i=1}^n \big(t_i,C^M t_i \big)}
e^{-\frac{1}{2} \sum_{i \neq j} \big(t_i,C^0 t_j \big)}
& \mbox{for} \; \sum_{j=1}^n q_j \; = \; 0 \\ \; & \; \\
0 & \mbox{for} \; \sum_{j=1}^n q_j \; \neq \; 0 \\
\end{array} \right. \; ,
\end{equation}
where
\begin{equation}
C^m \; := \; \Big(-\triangle + m^2 \Big)^{-1} \; , \; m = \mu,M 
\; \; \; \; ; \; \; \; \; 
C^0(x) \; := \; \Big(-\triangle \Big)^{-1} (x) \; = \; 
-\frac{1}{4\pi}\left( \ln(x^2) + 2\gamma - \ln4 \right) \; ,
\end{equation}
and
$ q_j \; := \; \int \; d^2x \; t_j(x)$. 
In order to obtain a nonvanishing $\rho_B$ the neutrality condition implies 
\begin{equation}
\sum_{b=1}^N U_{Ib} (n_b - m_b) \; \stackrel{!}{=} \; 0 \; \; \; \; \; \; 
\forall \; \; I = 2,3, ... N  \; \; .
\end{equation}
Interpreting the lines of $U_{Ib}$
as vectors $\vec{r}^{\;(I)}$ (see (68) for the definition of $U$),
the condition reads
\begin{equation}
(\vec{n} - \vec{m}) \cdot \vec{r}^{\;(I)} \; \stackrel{!}{=} \; 0 \; \; \; \; 
\forall \; \; I = 2,3, ... N  \; \; .
\end{equation}
One finds that the only solution is 
\begin{equation}
\vec{n} - \vec{m} \; \; \propto \; \; (1,1, \; .... \; 1) \; .
\end{equation}
Since $n_b$ and $m_b$ are integers this solution is equivalent to 
multiplication with $h(n_b,m_b)$. Thus the neutrality condition 
is the mechanism on the bosonic side which reproduces the selection rule
steming from the definition (51) of the clustering expectation
functional $\langle .. \rangle^\theta_0$.

A straightforward but lengthy computation shows that the equality
(90) can be fulfilled by setting the constants $c^{(b)}$ in (87) to
\begin{equation}
c^{(b)} = \Big(\frac{M^{(1)} e^\gamma}{2} 
\Big)^{\frac{\pi}{\pi+gN} \frac{1}{N}} \; 
\prod_{I=2}^N \Big(\frac{M^{(I)} e^\gamma}{2} \Big)^{(U_{Ib})^2} \; .
\end{equation}
Thus the bosonization is given by (86) and (87) together with (97).
The bosonic model describes the physical sector of the currents and the
chiral densities.

Having at hand the bosonization, one can immediately draw a second lesson
on the structure of the vacuum functional $\langle .. \rangle^\theta_0$.
The vacuum structure manifests itself in symmetry properties of the 
bosonized model.
\vskip5mm
\noindent
{\bf Lesson 2 :} (On the U(1)-Problem of QED$_2$.)
\vskip3mm
\noindent 
{\it The axial U(1)-symmetry is not a symmetry on the physical
Hilbert space, and there is no  U(1)-problem for QED$_2$.}
\vskip3mm
\noindent
This can be seen rather easily in the $\mbox{N}=2$ flavor case. The Lagrangian
for the scalar fields $\Phi^{(1)},\; \Phi^{(2)}$ that bosonize
the currents and the chiral densities then, is given by
\begin{equation}
\frac{1}{2} \Big( \partial_\mu \Phi^{(1)} \Big)^2 \; + \; 
\frac{1}{2} \Big( \partial_\mu \Phi^{(2)} \Big)^2 \; + \;
\frac{1}{2} \Big( \Phi^{(1)} \Big)^2 \frac{2 e^2 }{\pi+2g} \; .
\end{equation}
The bosonization prescription (87) for the left-handed densities gives
\[
\overline{\psi}^{(1)}(x) P_+ \psi^{(1)}(x) \; \longleftrightarrow \;
\frac{1}{2\pi} c^{(1)} 
:e^{-ia \Phi^{(1)}(x)}:_{M^{(1)}}
:e^{-ib \Phi^{(2)}(x)}:_{M^{(2)}} \; e^{i\frac{\theta}{2}} \; , \]
\begin{equation}
\overline{\psi}^{(2)}(x) P_+ \psi^{(2)}(x) \; \longleftrightarrow \;
\frac{1}{2\pi} c^{(2)} 
:e^{-ia \Phi^{(1)}(x)}:_{M^{(1)}}
:e^{+ib \Phi^{(2)}(x)}:_{M^{(2)}} \; e^{i\frac{\theta}{2}} \; ,
\end{equation}
with $a = \sqrt{2} \pi /\sqrt{\pi+2g}$ and
$b = \sqrt{2 \pi}$. The axial transformation (49) acts on the 
densities via
\begin{equation}
\overline{\psi}^{(b)}(x) P_+ \psi^{(b)}(x) \; \longrightarrow \; 
\overline{\psi}^{(b)}(x) P_+ \psi^{(b)}(x) \; e^{i 2 \omega} \; .
\end{equation}
In the bosonized theory this transformation corresponds to
\begin{equation}
a \; \Phi^{(1)}(x) \; + \; b \Phi^{(2)}(x) \; \longrightarrow \; 
a \; \Phi^{(1)}(x) \; + \; b \Phi^{(2)}(x) \; - \; 2 \omega \; ,
\end{equation}
and
\begin{equation}
a \; \Phi^{(1)}(x) \; - \; b \Phi^{(2)}(x) \; \longrightarrow \; 
a \; \Phi^{(1)}(x) \; - \; b \Phi^{(2)}(x) \; - \; 2 \omega \; .
\end{equation}
Obviously this is not a symmetry, since $\omega$ on the right hand sides of
(101), (102) cannot be transformed away, by shifting one of the fields 
$\Phi^{(1)}, \Phi^{(2)}$ by a
constant. $\Phi^{(1)}$ cannot be shifted since it is
massive (see (98)). $\Phi^{(2)}$ would have to be shifted by
$+2 \omega/b$ in order to remove $\omega$ in (101) and by
$-2 \omega/b$ to remove it in (102). Thus $\mbox{U(1)}_A$
is not a symmetry. The generalization
of the arguments to $\mbox{N} > 2$ flavors is straightforward.

What this lesson could tell us for QCD, is the simple statement
that the U(1)-problem does not exist. 
The charge $\tilde{Q}_5$ which formally \cite{bardeen} 
is supposed to generate $U(1)_A$, is not gauge invariant. 
Thus it has to be doubted \cite{lopu}, that U(1)$_A$ is a symmetry on the
physical Hilbert space (maybe it is a symmetry on a `larger space').
Hence the Goldstone theorem does not apply to the 
physical sector and there is no reason to expect a physical 
Goldstone particle. 
A proof that the unphysical sector decouples from physical
amplitudes is of course much more difficult in QCD, than the
simple arguments given for QED$_2$ above.

%
%
\subsection{The massive model and the GSG}
In the last subsection it was shown that it is possible
to find a common
bosonization of the Cartan currents together with
the chiral densities. In the mass perturbation series
(12) the expansion of the mass term of the action leads to
insertions of fermion mass term and thus to insertions of 
the chiral densities. Thus one can formally identify an
interaction term $S_{int}$ for the scalar fields which corresponds to the
mass term (9). By inserting the bosonization prescription (87) into
(9) one finds
\[
S_{int} [ \Phi^{(I)} ] \; := \; 
- \frac{1}{\pi} \sum_{b=1}^N m^{(b)} c^{(b)} \int d^2 x \; \chi_\Lambda(x)
\]
\begin{equation}
: \cos \left( 
2 \sqrt{\pi} \sqrt{\frac{\pi}{\pi\!+\!gN}} U_{1 b} \Phi^{(1)}(x)
+ 2\sqrt{\pi} \sum_{I=2}^N U_{I b} \Phi^{(I)}(x) - \frac{\theta}{N} \right) : 
\; .
\end{equation} 
The Wick ordering of the cosine is understood in terms of the 
perturbation expansion and thus reduces to the Wick ordering of 
exponentials (compare (87)). Note that the infrared cutoff $\Lambda$ 
is taken over. Also the role of the Thirring term which manifests
itself in a nonvanishing $g$ shows up in a new light. It leads
to the extra factor $\sqrt{\pi/(\pi\!+\!gN)}$ attached to the field 
$\Phi^{(1)}$ in the cosine, which keeps the model below the first
collapse point (see e.g. \cite{collapse}). 

In terms of the bosonized model the perturbation series (12) reads
\begin{equation}
\langle P[\{\Phi^{(I)}\}] \rangle \;  =  \; 
\frac{1}{Z} 
\sum_{n=0}^\infty \frac{(-1)^n}{n!}
\; \langle \; P[\{\Phi^{(I)}\}] \; S_{int}[ \{\Phi^{(I)}\}]^n
\; \rangle_Q \; .
\end{equation}
Using the bosonized model in \cite{diss1}, \cite{singo} it is proven 
that the perturbation series converges for small fermion masses $m^{(b)}$
and space-time cutoff $\Lambda$. Using the known techniques
for $N>1$ it is not possible to remove the cutoff termwise
due to the massless fields $\Phi^{(I)}, \; I > 1$ 
which show up for more than one flavor. For $N=1$ no such problem is faced
\cite{seilerfroh}.
Nevertheless for finite $\Lambda$ the bosonization is rigorous.
Only the 
perturbative treatment (104) breaks down for $\Lambda \rightarrow \infty$, 
a fact that is related to
logarithmic contributions in the small-mass behaviour of the fermion
determinant in infinite volume \cite{diss1}. For further analysis of this
non-perturbative behaviour from a different point of view see also 
\cite{colemanflav}, \cite{hosotani1,hosotani2}.

The bosonization to the model with the interaction (103) is a simple
generalization of the Coleman isomorphism which maps the 
one-flavor Schwinger model to the Sine-Gordon model 
\cite{colemaniso1, colemaniso2}.
Models of the type (103) but without the UV-regulator were already 
discussed in \cite{belvedere}. 
The classical Lagrangian ${\cal L}_{GSG}$
which corresponds to the generalized Sine-Gordon model can be read off
from the $Q^{(I)}$ (see (84), (85)) and $S_{int}$
\[
{\cal L}_{GSG} \; = \; 
\frac{1}{2} \sum_{I=1}^N \partial_\mu \Phi^{(I)} \partial_\mu \Phi^{(I)} 
\; + \; \frac{1}{2} \Big( \Phi^{(1)} \Big)^2 \; \frac{e^2 N}{\pi+gN}
\]
\begin{equation}
- \; \frac{1}{\pi} \sum_{b=1}^N m^{(b)} c^{(b)} 
\cos \left( 
2 \sqrt{\pi} \sqrt{\frac{\pi}{\pi\!+\!gN}} U_{1 b} \Phi^{(1)}
+ 2\sqrt{\pi} \sum_{I=2}^N U_{I b} \Phi^{(I)} - \frac{\theta}{N} \right) 
\; .
\end{equation} 
It has to be remarked that the GSG defined through (105) bosonizes 
only the $N$ Cartan currents (75) together with the chiral densities,
although there are all together $N^2$ vector currents in the $N$-flavor model.
In \cite{seilergatt} it was shown that there is no common abelian 
bosonization of all $N^2$ vector currents. However only the U(1) current
$J^{(1)}_\mu$
plays a special role. It gives rise to a heavy state, while the other $N^2 -1$
states remain light. $N-1$ of the currents corresponding to light states
are now bosonized together with the U(1) current. This is sufficient for the
discussion below.

The explicit form (105) of the Lagrangian of the GSG
allows to draw another lesson that recovers a property
of the $\theta$-vacuum in QCD.
\vskip3mm
\noindent
{\bf Lesson 3 :} (On the vacuum structure of QED$_2$.)
\vskip3mm
\noindent
{\it Physics does not depend on $\theta$ if at least one of the 
fermion masses vanishes.}
\vskip3mm
\noindent
This property can be seen to hold in the bosonized version
by the following arguments
\begin{equation}
U_{N 1} \; = \; \frac{1}{\sqrt{2}} \; \; \; , \; \; \; 
U_{N 2} \; = \; \frac{-1}{\sqrt{2}} \; \; \; , \; \; \;
U_{N b} \; = \; 0 \; \; \; \; \mbox{for} \; \; 3 \leq b \leq N \; \; ,
\end{equation}
for $N \geq 2$ (compare (68)), one obtains for the interaction term in (105)
\[
\sum_{b=1}^N m^{(b)} c^{(b)} 
: \cos \left( 
2 \sqrt{\pi} \sqrt{\frac{\pi}{\pi\!+\!gN}} U_{1 b} \Phi^{(1)}(x)
+ 2\sqrt{\pi} \sum_{I=2}^N U_{I b} \Phi^{(I)}(x) - \frac{\theta}{N} \right) : 
\]
\[ 
= \; m^{(1)} c^{(1)} : \cos \Bigg( 
2 \sqrt{\pi} \sqrt{\frac{\pi}{\pi\!+\!gN}} U_{1 b} \Phi^{(1)}(x)
+ 2\sqrt{\pi} \sum_{I=2}^{N-1} U_{I b} \Phi^{(I)}(x) 
\]
\[
+ \; 2\sqrt{\pi} \frac{1}{\sqrt{2}}\Phi^{(N)}(x)
- \frac{\theta}{N} \Bigg) :
\]
\begin{equation}
+ \; 
\sum_{b=3}^N m^{(b)} c^{(b)} 
: \cos \left( 
2 \sqrt{\pi} \sqrt{\frac{\pi}{\pi\!+\!gN}} U_{1 b} \Phi^{(1)}(x)
+ 2\sqrt{\pi} \sum_{I=2}^{N-1} U_{I b} \Phi^{(I)}(x) -
\frac{\theta}{N} \right) :
\; .
\end{equation}
Since $\Phi^{(N)}$ is a massless field and shows up only in
the first term on the right hand side of (107) it can be shifted by a constant
in order to change $\theta$. If none of the masses vanishes, $\Phi^{(N)}$
enters the interaction term twice but with different signs, as can
be seen from (68). The value of $\theta$ cannot be changed by a 
symmetry transformation then, and 
physics depends on it.

The above property is believed to hold also for the formal
$\theta$-vacuum of QCD \cite{callan1,callan2}. The independence of $\theta$ 
can be seen also from an 
alternative introduction of the vacuum angle \cite{masstheta1,masstheta2}. 
It is the starting point for the derivation of the Witten-Veneziano type
formulas \cite{witten1}-\cite{smit}, \cite{seilerstam}.

%
%
\subsection{Semiclassical approximation and Witten-Veneziano type formulas}
Since in the perturbative treatment the space-time cutoff $\Lambda$ 
spoils Lorentz invariance and thus the extraction of physical quantities,
one is reduced to a semiclassical approximation of the Lagrangian.

In order to simplify the involved structure of the interaction (105) we
consider the special case of all fermion masses being equal
$m^{(b)} := m$ for $b=1,2 \; .. \; N$. Using the fact that in
${\cal L}_{GSG}$ only $\Phi^{(1)}$ plays an extra role one can
furthermore restrict the masses $M^{(I)}$ used for normal ordering to
$M^{(I)} = M$ for $I=2,3 \; .. \; N$. Inserting this restriction in 
formula (97) for the coefficients $c^{(b)}$ showing up in ${\cal L}_{GSG}$
one finds $c^{(b)} := c$ for $b=1,2 \; .. \; N$. Together with the 
restriction for the fermion masses this reduces the semiclassical problem 
to linear algebra and to the solution of only one transcendental
equation. Without this restriction one would have to solve a system of 
$N$ coupled transcendental equations.

The minima $\Phi^{(I)}_0$ of the potential $V( \Phi^{(I)} )$ have to be 
computed. The potential $V( \Phi^{(I)} )$ is given by
\begin{equation}
V( \Phi^{(I)} ) \; := \; 
\frac{1}{2} \Big( \Phi^{(1)} \Big)^2 \; \frac{e^2 N}{\pi+gN} \; - \;
\frac{1}{\pi} m c \sum_{b=1}^N 
\cos \left( 
2 \sqrt{\frac{\pi}{N}} \sqrt{\frac{\pi}{\pi\!+\!gN}} \Phi^{(1)}
+ 2\sqrt{\pi} \sum_{I=2}^N U_{I b} \Phi^{(I)} - \frac{\theta}{N} \right) 
\; .
\end{equation}
Setting $\frac{\partial}{\partial \Phi^{(J)}} 
V( \Phi^{(I)} ) \big|_{\Phi^{(I)} = \Phi^{(I)}_0} = 0$ gives
\begin{equation}
\frac{e^2 N}{\pi\!+\!gN} \Phi^{(1)}_0 + \frac{2mc}{\sqrt{\pi\!+\!gN}}
\sum_{b=1}^N \frac{1}{\sqrt{N}}
\sin \left( 2 \sqrt{\frac{\pi}{N}} 
\sqrt{\frac{\pi}{\pi\!+\!gN}} \Phi^{(1)}_0
+ 2\sqrt{\pi} \sum_{I=2}^N U_{I b} \Phi^{(I)}_0 - \frac{\theta}{N} \right) 
\; = \; 0 \; ,
\end{equation}
for $J=1$ and
\begin{equation}
\frac{2mc}{\sqrt{\pi}} \sum_{b=1}^N U_{J b}
\sin \left( 2\sqrt{\frac{\pi}{N}} \sqrt{\frac{\pi}{\pi\!+\!gN}} \Phi^{(1)}_0
+ 2\sqrt{\pi} \sum_{I=2}^N U_{I b} \Phi^{(I)}_0 - \frac{\theta}{N} \right) 
\; = 0 \; 
\; ,
\end{equation}
for $J = 2,3, \; ... \;N$.
Again one can interpret the lines of $U$ (fixed $J$ in $U_{Jb}$ ) as
vectors $\vec{r}^{\;(J)}$ (compare (68)) and denote
Eq. (110) as products of two vectors
\begin{equation}
\vec{r}^{\;(J)} \cdot \vec{s} \; \stackrel{!}{=} \; 0 \; \; \; \; \; \; \; \; 
\forall \; \; J \; = \; 2,3\; ... \; N \; ,
\end{equation}
where the entries of the vector $\vec{s}$ are given by
\begin{equation}
s_b := 
\frac{2mc}{\sqrt{\pi}} \sin \left( 2 \sqrt{\pi} \sqrt{\frac{\pi}{\pi\!+\!gN}} 
\frac{1}{\sqrt{N}} \Phi^{(1)}_0
+ 2\sqrt{\pi} \sum_{I=2}^N U_{I b} \Phi^{(I)}_0 - \frac{\theta}{N} \right) 
\; .
\end{equation}
We already found (compare (95),(96)) that the only solution is 
$\vec{s} \; = \; \lambda \; ( 1,1, \; ... \; 1 )\; , \;
\lambda \; \in \; I\!\!R$ .
Thus the set of Eqs. (110) is equivalent to
\begin{equation}
2 \sqrt{\pi} \sqrt{\frac{\pi}{\pi\!+\!gN}} 
\frac{1}{\sqrt{N}} \Phi^{(1)}_0
+ 2\sqrt{\pi} \sum_{I=2}^N U_{I b} \Phi^{(I)}_0 - \frac{\theta}{N}  \; = \; 
\arcsin \left( \frac{ \lambda \sqrt{\pi}}{2m c} \right) \; ,
\end{equation}
for all $b = 1,2, ... N$. 
Eq. (109) can now be used to express $\lambda$ in terms of $\Phi^{(1)}_0$ 
giving
$\lambda  = - \Phi^{(1)}_0 \sqrt{N} e^2 / \sqrt{\pi(\pi + gN)} 
\Phi^{(1)}_0$.
Inserting this in (113), multiplying with $U_{Jb}$ and summing over $b$ gives
\begin{equation}
\sum_{I=2}^N \delta_{JI} \Phi^{(I)}_0 = 
\delta_{J1} \left\{\frac{\sqrt{N}}{2\sqrt{\pi}}
\left[\frac{\theta}{N} - \arcsin\left( 
\frac{e^2}{2mc} \sqrt{\frac{N}{\pi\!+\!gN}}  
\Phi^{(1)}_0 \right) \right] - \sqrt{\frac{\pi}{\pi\!+\!gN}} \Phi^{(1)}_0  
\right\} \; ,
\end{equation}
where we used the orthogonality of $U$ and $\sum_b U_{Jb} = \delta_{J1} 
\sqrt{N}$ (see(68)). 
In the last expression the equations for the determination 
of the minima are decoupled and can be solved easily. 
For the case $2 \leq J \leq N$ one obtains
the naive solution
$\Phi_0^{(J)} \; = \; 0$ for $J=2,3 \; .. \; N$.
Of course there exists an infinite countable set of solutions 
since one can always shift the argument of the cosine in (108)
by integer multiples
of $2\pi$
giving rise to
\begin{equation}
2 \sqrt{\pi} \sum_{J=2}^N U_{J b} \Phi^{(J)}_0 = n_b 2 \pi \; ,\; \; \; \; 
n_b \; \in \; \mbox{Z\hspace{-1.35mm}Z} 
\; \; ,\; \; \; \; \forall \; \; b \; = 
\; 1,2, \; ... \; N \; .
\end{equation}
Using the orthogonality of $U$, 
the last expression can be inverted and one ends up with
\begin{equation}
\Phi^{(I)}_0 \; = \; \sqrt{\pi} \sum_{b=1}^N U_{Ib} \; n_b \; \; \; \; \; 
\; I \; = \; 2,3,\; ... \; N \; .
\end{equation}
The $\Phi^{(1)}_0$ coordinate of the minimum has to fulfill the
equation that emerges from (114) setting $J=1$
\begin{equation}
\frac{1}{\sqrt{N}}\sqrt{\frac{\pi}{\pi\!+\!gN}} \Phi^{(1)}_0
\; = \;
\frac{1}{2\sqrt{\pi}}
\left[ \frac{\theta}{N} - 
\arcsin\left( \frac{e^2}{2mc} \sqrt{\frac{N}{\pi\!+\!gN}}  
\Phi^{(1)}_0 \right) \right] \; .
\end{equation}
Obviously this is a trivial modification of the transcendental equation
that determines the minimum in the one flavor case. It has to be solved 
numerically. 

To evaluate the mass matrix of the effective theory around the minima,
the Hesse matrix 
$H_{I I^\prime} \; := \; \frac{ \partial^2 \; V (\Phi^{(J)})}
{\partial \Phi^{(I)} \partial \Phi^{(I^\prime)}} \; \; 
\big|_{\Phi^{(J)} = \Phi^{(J)}_0}$ has to be computed.
It can be evaluated easily
\[
H \; = \; \mbox{diag} \Big( \frac{e^2 N}{\pi\!+\!gN}, 0 , \; ... \; 0 \Big)
\; + 
\]
\[
4 mc \; \tilde{\lambda} \; \sum_{b=1}^N \left[
\begin{array}{cccccc}
\frac{\pi}{\pi\!+\!gN} \frac{1}{N} & 
\sqrt{\frac{\pi}{\pi\!+\!gN}} \sqrt{\frac{1}{N}} U_{2b} & . & . & . &
\sqrt{\frac{\pi}{\pi\!+\!gN}} \sqrt{\frac{1}{N}} U_{Nb} \\
\sqrt{\frac{\pi}{\pi\!+\!gN}} \sqrt{\frac{1}{N}} U_{2b} &
U_{2b} U_{2b} & . & . & . & 
U_{2b} U_{Nb} \\
\sqrt{\frac{\pi}{\pi\!+\!gN}} \sqrt{\frac{1}{N}} U_{3b} &
U_{3b} U_{2b} & \; & \; & \; & \; \\
. & . & . & \; & \; & . \\
. & . & \; & . & \; & . \\
. & . & \; & \; & . & . \\
\sqrt{\frac{\pi}{\pi\!+\!gN}} \sqrt{\frac{1}{N}} U_{Nb} &
U_{Nb} U_{2b} & . & . & . & U_{Nb} U_{Nb}
\end{array} \right] \; = 
\] 
\\
\begin{equation}
\mbox{diag} \Big( \frac{e^2 N}{\pi\!+\!gN} 
+ \frac{\pi}{\pi\!+\!gN} 4 mc \tilde{\lambda} \; , \;  4 mc \tilde{\lambda} \;
, \; . \; . \; . \; .\; . \; , \; 
4 mc \tilde{\lambda} \Big) \; .
\end{equation}
The orthogonality of $U$ was used again.
$\tilde{\lambda}$ is defined as 
\begin{equation}
\tilde{\lambda} \; := \; \cos \arcsin 
\left( \frac{\lambda \sqrt{\pi}}{2mc} \right) \; = \; 
\sqrt{1- \left( \frac{\lambda \sqrt{\pi}}{2mc} \right)^2} \; .
\end{equation}
The Hesse matrix comes out as a positive definite diagonal matrix. The
entries have to be interpreted as the squared masses of the fields $\Phi^{(I)}$
in an effective theory around the semiclassical vacua. The masses $m_I$ 
for the fields $\Phi^{(I)}$ are given by
\begin{equation}
m_1 \; := \; \sqrt{  \; \frac{e^2 N}{\pi} + 4 mc \tilde{\lambda}  } \; 
\sqrt{ \frac{\pi}{\pi\!+\!gN}} \; \; \; \; \; \; \; \mbox{for} \; \; \; 
\Phi^{(1)} \; ,
\end{equation}
and
\begin{equation}
m_I \; := \; \sqrt{ 4 mc \tilde{\lambda} } \; 
\; \; \; \; \; \; \; \mbox{for} \; \; \; 
\Phi^{(I)}\; , \; \; \; I \; = \; 2,3, \; ... \; N \; .
\end{equation}
It is interesting to notice that the masses $m_I$ do not depend
on $\theta$. It also has to be emphasized that the semiclassical approximation 
of the GSG has two regimes where the approximation is good. Firstly this is
the case for large fermion masses  $m$ which is the usual domain of a 
semiclassical approximation. Secondly 
for $m \rightarrow 0$ this 
approximation becomes exact, since the massless model is bosonized by 
free fields (compare (71)) where the semiclassical approximation already 
gives the spectrum of the quantized theory. We remark that the 
linear behaviour (120), (121) in $m$, is modified in the small $m$ regime
for $N>1$. However for the case of 
$\theta = 0$, the semiclassical approximation can be 
computed in closed form without solving a transcendental equation. 
For the $N=1$ case one then
finds that the semiclassical result then coincides with the first order 
result \cite{adam1,adam2}, of the perturbation expansion in $m$, 
which has a sound basis for $N=1$ \cite{seilerfroh}.

The masses obtained in the semiclassical 
approximation will now be used to test Witten-Veneziano formulas.
Since the semiclassical arguments do not rely on finite $g$, 
we set $g$ to zero in the following. 
Of course one could modify the 
Witten-Veneziano formula (122) (see below)
to include a finite $g$. 

For $g=0$ the following generalization of the Witten-Veneziano formula 
quoted in \cite{seilergatt} will be shown to hold:
\begin{equation}
m_1^2 \; - \; 
\frac{1}{N-1} \sum_{I=2}^N \;m^2_I \; = \; \frac{4N}{(f^0_1)^2} \; P^0(0) \; .
\end{equation}
$f^0_1$ denotes the decay constant of the U(1)-pseudoscalar, and $P^0(0)$
is the contact term of the topological susceptibility \cite{seilerstam} 
defined through the spectral representation 
(see also \cite{diss1, seilergatt})
\begin{equation}
\frac{e^2}{(2\pi)^2}
\int \langle F_{12}(x) F_{12}(0) \rangle e^{-ipx}dx \; \;  =  \; \; 
P^0(0) - \int_0^\infty \frac{d\rho(\mu^2)}{p^2 + \mu^2} \; .
\end{equation}
Inserting the mass values (120) and (121) at $g=0$, one
finds that the left hand side of (122) reduces to
\begin{equation}
m_1^2 \; - \; 
\frac{1}{N-1} \sum_{I=2}^N \;m^2_I \; \; = \; \; \frac{e^2 N}{\pi} \; .
\end{equation}
In \cite{seilergatt} $P^0(0)$ and $f^0_1$ were 
computed explicitly, and it was shown that the right hand side of (124) 
can be rewritten as the right hand side of (122) and thus (122) is 
proven.
This result is our Lesson 4.
\newpage
\noindent
{\bf Lesson 4 :} (A Witten-Veneziano-type formula for QED$_2$.)
\vskip3mm
\noindent
{\it The masses (determined from a semiclassical approximation) 
of the particles that correspond to the Cartan currents
obey the Witten-Veneziano formula} (122).
\vskip3mm
\noindent
It has to be remarked that (122) is also
a verification of the original form of the Witten-Veneziano formula,
because the topological susceptibility of the quenched theory reduces
to the contact term \cite{seilerstam, seilergatt}. 
It is not true, however, that the topological
susceptibility appearing in the formula expresses a property of the long
distance fluctuations of the topological density. In fact it is entirely
given by the contact term expressing short distance fluctuations.

%
%
\section{Discussion}
When trying to take over part of the Lessons 1-4 
for QED$_2$ to the case of QCD, one of course 
never should forget the limitations of a two dimensional toy model. 

As was discussed, the Coleman theorem \cite{colemangold} determines the
form of the chiral selection rule (57) quoted in Lesson 1 for QED$_2$. 
On the other hand Coleman's theorem does not allow for
\begin{equation}
\langle \; \theta \; | \; \overline{q} \; q \; | \; 
\theta \; \rangle \; \neq 0 \; \; ,
\end{equation}
in QED$_2$ with more than one flavor.
For QCD Eq. (125) is one of the main assumptions
supporting the belief that QCD is the correct theory for
strong interactions. Thus Coleman's theorem limits the relevance of 
Lesson 1 for QCD. 

What Lesson 2 tries to tell us is of more direct relevance for QCD.
The charge \cite{bardeen} that is used to formally implement the U(1)-axial
symmetry is not gauge invariant. Thus it has to be questioned
if the U(1)$_A$ symmetry can be implemented on the physical 
Hilbert space. If not, the Goldstone theorem does not apply,
and there is no reason to expect a light pseudoscalar. 
Thus QED$_2$ indeed suggests the
most simple solution to the U(1)-problem: 
The U(1)-problem does not exist. As already discussed, the proof 
that unphysical particles decouple from the physical spectrum is
much more involved for QCD.

Lesson 3 recovers a property that is commonly accepted to hold for the 
$\theta$-vacuum of QCD. 

The main limitation of taking over Lesson 4 to QCD, is that 
the topological susceptibility of QED$_2$ is too simple 
to model the problems that show up in QCD. In particular $\chi_{top}$
of QCD is a composite operator that requires some subtraction 
procedure determining its properties.
As a further limitation it has been pointed
out \cite{shif} that the structure of the SU(N)-currents  
that play the role of the pseudoscalar mesons is too simple.
In particular they obey the current algebra of free fermions
\cite{halpern1,halpern2} and thus their interaction is reduced to flavor
exchange.

Despite the limitations of the lessons for QCD, the study of 
the corresponding problems in QED$_2$ is interesting, in particular 
because 
techniques independent from the concepts used for QCD were developed.
For QED$_2$ it has been demonstrated that it is 
possible to define a clustering vacuum functional without making 
use of formal instanton arguments.
We believe that this is the conceptually clearer way for a two dimensional
model where the mathematical structure is rather simple. 
\vskip3mm
\noindent
The author acknowledges intensive scientific support by Erhard 
Seiler from Max-Planck-Institute in Munich, interesting discussions with 
Peter Forgacs from the University of Tours
and thanks Rod Crewther for correspondence.

\newpage
%
%


\begin{thebibliography}{1234567} 
\newcommand{\bibi}[1]{\bibitem{#1}}
\newcommand{\authors}[1]{#1, }
\newcommand{\journal}[1]{#1}
\newcommand{\volume}[1]{{\bf #1}}
\newcommand{\myyear}[1]{(#1)}
\newcommand{\page}[1]{#1}
\bibi{schwinger}
\authors{J. Schwinger}
\journal{Phys. Rev.} 
\volume{128} \myyear{1962} \page{2425}.

\bibi{callan1}
\authors{C.G. Callan, R.F. Dashen and D.J. Gross}
\journal{Phys. Lett.} 
\volume{B63} \myyear{1976} \page{334}.

\bibi{callan2} 
\authors{R. Jackiw and C. Rebbi}
\journal{Phys. Rev. Lett.}
\volume{37} \myyear{1976} \page{172}.

\bibi{lowenstein}
\authors{J.H. Lowenstein and J.A. Swieca}
\journal{Ann. Phys.} \volume{68} \myyear{1971} \page{175}.

\bibi{jack2}
\authors{R. Jackiw}
\journal{Rev. Mod. Phys.} \volume{52} \myyear{1980} \page{661}.

\bibi{colella}
\authors{P. Colella and O.E. Lanford III}
\journal{in
"Constructive Quantum Field Theory (Erice 1973)" (G. Velo, A. Wightman, 
Eds.),
Springer Lecture Notes in Physics 25, New York, 1973}.

\bibi{joos1}
\authors{H.Joos}
\journal{Helv. Phys. Acta} 
\volume{63} \myyear{1990} \page{670}.

\bibi{joos2}
\authors{H.Joos and S.I. Azakov}
\journal{Helv. Phys. Acta} 
\volume{67} \myyear{1994} \page{723}.

\bibi{wipf1}
\authors{I. Sachs and A. Wipf}
\journal{Helv. Phys. Acta} 
\volume{65} \myyear{1992} \page{653}.

\bibi{wipf2}
\authors{I. Sachs, A. Wipf and A. Dettki}
\journal{Phys. Lett.} 
\volume{B317} \myyear{1993} \page{545}.

\bibi{wipf3}
\authors{A. Dettki, I. Sachs and A. Wipf}
\volume{hep-th/9308067}.

\bibi{wipf4}
\authors{I. Sachs and A. Wipf}
\journal{Phys. Lett.} 
\volume{B326} \myyear{1994} \page{105}.

\bibi{wipf5}
\authors{I. Sachs}
\journal{"On Symmetries, their Breaking, and 
Thermodynamics in Field Theory" Thesis, ETH Z\"urich, 1994} 
\volume{Diss.ETH No. 10728}.

\bibi{wipf6}
\authors{A. Wipf and S. D\"urr}
\volume{hep-th/9412018}.

\bibi{weinberg}
\authors{S. Weinberg}
\journal{Phys. Rev.}
\volume{D11} \myyear{1975} \page{3583}.

\bibi{thooft1}
\authors{G. 't Hooft}
\journal{Phys. Rev. Lett.}
\volume{37} \myyear{1976} \page{8}.

\bibi{thooft2}
\authors{G. 't Hooft}
\journal{Phys. Rev.}
\volume{D14} \myyear{1976} \page{3432}.

\bibi{thooft3}
\authors{G. 't Hooft}
\journal{Phys. Rept.}
\volume{142} \myyear{1986} \page{357}.

\bibi{crewther1}
\authors{R.J. Crewther}
\journal{Phys. Lett.}
\volume{B70} \myyear{1977} \page{349}.

\bibi{crewther2}
\authors{G.A. Christos}
\journal{Phys. Rept.}
\volume{116} \myyear{1984} \page{751}. 

\bibi{bardeen}
\authors{W.A. Bardeen}
\journal{Nucl. Phys.}\volume{B75} 
\myyear{1974} \page{246}.

\bibi{seilerstam}
\authors{E.Seiler and I.O.Stamatescu}
\journal{Preprint (unpublished)} 
\volume{MPI-PAE/PTh 10/87} \myyear{1987}. 

\bibi{diss1}
\authors{C. Gattringer}
\journal{"$\mbox{QED}_2$ and U(1)-problem" Thesis, University of Graz, 1995,} 
\volume{hep-th/9503137}.

\bibi{diss2}
\authors{C. Gattringer}
\journal{in
"Low-Dimensional Models in Statistical Physics and Quantum Field Theory"
(Proceedings of
34. Internationale Universit\"atswochen f\"ur Kern- und Teilchenphysik, 
Schladming, 1995)}
\volume{hep-th/9505092}. 

\bibi{witten1}
\authors{E. Witten}
\journal{Nucl. Phys.}
\volume{B160} \myyear{1979} \page{57}.

\bibi{witten2}
\authors{G. Veneziano}
\journal{Nucl. Phys.}
\volume{B159} \myyear{1979} \page{213}.

\bibi{smit}
\authors{J. Smit and J.C. Vink}
\journal{Nucl. Phys.}
\volume{B284} \myyear{1987} \page{234}.

\bibi{weingarten1}
\authors{D.H. Weingarten and J.L. Challifour}
\journal{Ann. Phys.} 
\volume{123} \myyear{1979} \page{61}.

\bibi{weingarten2}
\authors{D.H. Weingarten}
\journal{Ann. Phys.} 
\volume{126} \myyear{1980} \page{154}.

\bibi{seiler}
\authors{E. Seiler}
\journal{"Gauge Theories as a 
Problem of Constructive Quantum Field Theory and
Statistical Mechanics" Springer Lecture notes in Physics 159, 
New York, 1982}.

\bibi{seilergatt}
\authors{C. Gattringer and E. Seiler}
\journal{Ann. Phys.} 
\volume{233} \myyear{1994} \page{97}. 

\bibi{deutsch}
\authors{C. Deutsch and M. Lavaud}
\journal{Phys. Rev.}
\volume{A9} \myyear{1974} \page{2598}.

\bibi{colemangold}
\authors{S. Coleman}
\journal{Comm. Math. Phys.}
\volume{31} \myyear{1973} \page{259}.

\bibi{rotheswi}
\authors{K.D. Rothe and J.A. Swieca}
\journal{Ann. Phys.}
\volume{117} \myyear{1979} \page{382}.

\bibi{belvedere}
\authors{L.V. Belvedere, J.A. Swieca, K.D. Rothe and B. Schroer}
\journal{Nucl. Phys.}
\volume{B153} \myyear{1979} \page{112}.

\bibi{smilga}
\authors{A.V. Smilga}
\journal{Phys. Lett.}
\volume{B278} \myyear{1992} \page{371}.

\bibi{peccei}
\authors{R. Peccei}
\journal{in "CP violation"
(C. Jarlskog Ed.), World Scientific, Singapore, 1989}.

\bibi{salm}
\authors{M. Salmhofer and E. Seiler}
\journal{Comm. Math. Phys.} 
\volume{139} \myyear{1991} \page{395}.

\bibi{crewtherkai}
\authors{R.J. Crewther}
\journal{ in
"Field Theoretical Methods in Particle Physics (NATO Summer School
1979 Kaiserslautern)" (W. R\"uhl, Ed.), Plenum Press, New York, 1980}.

\bibi{simonphi}
\authors{B. Simon}
\journal{"The P($\phi)_2$ Euclidean (Quantum) Field Theory"
Princeton University Press, Princeton, 1974}.

\bibi{frohlich1}
\authors{J. Fr\"ohlich}
\journal{
in "Renormalization Theorie (Erice 1975)",
(G. Velo, A.S. Wightman, Eds.), Reidel publishing company, 
Dordrecht, 1976}.

\bibi{lopu}
\authors{J. {\L}opusz{\'a}nski}
\journal{"An Introduction to Symmetry and Supersymmetry in
Quantum Field Theory" World Scientific, Singapore, 1991}. 

\bibi{collapse}
\authors{G. Benfatto, G. Gallavotti and F. Nicolo}
\journal{Comm. Math. Phys.} 
\volume{85} \myyear{1982} \page{387}.

\bibi{singo}
\authors{C. Gattringer}
\journal{"On the Construction of the N-Flavor Sine-Gordon Model"
(work in preparation)}.

\bibi{seilerfroh}
\authors{J. Fr\"ohlich and E. Seiler}
\journal{Helv. Phys. Acta}
\volume{49} \myyear{1976} \page{889}.

\bibi{colemanflav}
\authors{S. Coleman}
\journal{Ann. Phys.} \volume{101} \myyear{1976} \page{239}.

\bibi{hosotani1}
\authors{J.E. Hetrick, Y. Hosotani and S.Iso}
\volume{hep-th/9502113}.

\bibi{hosotani2}
\authors{Y. Hosotani}
\volume{hep-th/9505168}.

\bibi{colemaniso1}
\authors{S. Coleman}
\journal{Phys. Rev.} \volume{D11} \myyear{1975} \page{2088}.

\bibi{colemaniso2}
\authors{S. Coleman, R. Jackiw and L. Susskind}
\journal{Ann. Phys.} \volume{93} \myyear{1975} \page{267}.

\bibi{masstheta1}
\authors{V. Baluni}
\journal{Phys. Rev.}
\volume{D19} \myyear{1979} \page{2227}.

\bibi{masstheta2}
\authors{E. Seiler and I.O. Stamatescu}
\journal{Phys. Rev.}
\volume{D25} \myyear{1981} \page{2177}.

\bibi{shif}
\authors{M.A. Shifman and A.V. Smilga}
\journal{Phys. Rev.}
\volume{D50} \myyear{1994} \page{7659}.

\bibi{halpern1}
\authors{M. Halpern}
\journal{Phys. Rev.}
\volume{D12} \myyear{1975} \page{1684}.

\bibi{halpern2}
\authors{M. Halpern}
\journal{Phys. Rev.}
\volume{D13} \myyear{1976} \page{337}.

\bibi{adam1}
\authors{C. Adam}
\journal{preprint BUTP-95/27}
\volume{hep-ph 9507331}. 

\bibi{adam2}
\authors{J.P. Vary, T.J. Fields and H.J. Pirner}
\journal{preprint ISU-NP-94-14}
\volume{hep-ph 9411263}.

\end{thebibliography}
\end{document}